\documentclass[10pt,preprint]{aastex}

\usepackage{graphicx}
\usepackage{rotating}

\shorttitle{}
\shortauthors{Nesvorn\'y et al.}

\begin{document}
\baselineskip 19.pt

\title{NEOMOD 2: An Updated Model of Near-Earth Objects from a Decade of Catalina Sky Survey Observations}

\author{David Nesvorn\'y$^1$, David Vokrouhlick\'y$^2$, Frank Shelly$^3$, Rogerio Deienno$^1$,\\ 
William F. Bottke$^1$, Eric~Christensen$^3$, Robert Jedicke$^4$, Shantanu~Naidu$^5$, 
Steven~R.~Chesley$^5$,\\ Paul W. Chodas$^5$, Davide Farnocchia$^5$, Mikael Granvik$^{6,7}$ } 
\affil{(1) Department of Space Studies, Southwest Research Institute, 1050 Walnut St., 
  Suite 300,  Boulder, CO 80302, USA}
\affil{(2) Institute of Astronomy, Charles University, V Hole\v{s}ovi\v{c}k\'ach 2, CZ–18000 
Prague 8, Czech Republic}
\affil{(3) Lunar and Planetary Laboratory, The University of Arizona, 1629 E. University Blvd. 
Tucson, AZ 85721-0092, USA}
\affil{(4) Institute for Astronomy, University of Hawaii, 2680 Woodlawn Drive, Honolulu, HI
  96822-1839, USA}
\affil{(5) Jet Propulsion Laboratory, California Institute of Technology, 4800 Oak Grove Dr.
Pasadena, CA 91109, USA}
\affil{(6) Department of Physics, University of Helsinki, P.O. Box 64, FI-00014, Finland} 
\affil{(7) Asteroid Engineering Laboratory, Lule\aa{} University of Technology, Box 848, 
SE-981 28, Kiruna, Sweden}

\begin{abstract}
Catalina Sky Survey (CSS) is a major survey of Near-Earth Objects (NEOs). In a recent work, we used
CSS observations from 2005-2012 to develop a new population model of NEOs (NEOMOD). CSS's G96
telescope was upgraded in 2016 and detected over 10,000 unique NEOs since then. Here we characterize
the NEO detection efficiency of G96 and use G96's NEO detections from 2013-2022 to update NEOMOD.
This resolves previous model inconsistencies related to the population of large NEOs. We estimate there are
$936 \pm 29$ NEOs with absolute magnitude $H<17.75$ (diameter $D>1$ km for the reference albedo
$p_{\rm V}=0.14$) and semimajor axis $a<4.2$ au. The slope of the NEO size distribution for $H=25$-28 is
found to be relatively shallow (cumulative index $\simeq 2.6$) and the number of $H<28$ NEOs ($D>9$ m
for $p_{\rm V}=0.14$) is determined to be $(1.20 \pm 0.04) \times 10^7$, about 3 times lower than 
in Harris \& Chodas (2021). Small NEOs have a different orbital distribution and higher impact probabilities 
than large NEOs. We estimate $0.034 \pm 0.002$ impacts of $H<28$ NEOs on the Earth per year,
which is near the low end of the impact flux range inferred from atmospheric bolide observations. 
Relative to a model where all NEOs are delivered directly from the main belt, the population of 
small NEOs detected by G96 shows an excess of low-eccentricity orbits with 
$a \simeq 1$--1.6 au that appears to increase with $H$ ($\simeq 30$\% excess for $H = 28$). We suggest 
that the population of very small NEOs is boosted by tidal disruption of large NEOs during close 
encounters to the terrestrial planets. When the effect of tidal disruption is (approximately) accounted 
for in the model, we estimate $0.06 \pm 0.01$ impacts of $H<28$ NEOs on the Earth per year, which is 
more in line with the bolide data. The impact probability of a $H<22$ ($D>140$ m for $p_{\rm V}=0.14$) 
object on the Earth in this millennium is estimated to be $\simeq 4.5$\%.   

\end{abstract}

\section{Introduction}

NEOMOD is an orbital and absolute magnitude model of NEOs (Nesvorn\'y et al. 2023; hereafter 
Paper I). To develop NEOMOD, we closely followed the methodology from previous studies (Bottke et 
al. 2002, Granvik et al. 2018), and improved it when possible. First, massive numerical 
integrations were performed for asteroid orbits 
escaping from eleven main belt sources. Comets were included as the twelfth source. The integrations 
were used to compute the probability density functions (PDFs) that define the orbital distribution 
of NEOs (perihelion distance $q<1.3$ au, $a<4.2$ au) from each source. Second, we developed a new method 
to accurately calculate biases of NEO surveys and applied it to the Catalina Sky Survey (CSS; Christensen 
et al. 2012) in an extended magnitude range ($15<H<28$). The publicly available 
\texttt{objectsInField}\footnote{\url{https://github.com/AsteroidSurveySimulator/objectsInField}}
code ({\texttt{oIF}) from the Asteroid Survey Simulator (AstSim) package (Naidu et al. 2017)
was used to to determine the geometric bias of CSS. Third, we used the \texttt{MultiNest} 
code, a Bayesian inference tool designed to efficiently search for best-fitting solutions in high-dimensional 
parameter space (Feroz \& Hobson 2008, Feroz et al. 2009), to optimize the biased model fit to CSS 
detections. The final model was made available to the scientific community via a NEOMOD 
Simulator\footnote{\url{https://www.boulder.swri.edu/\~{}davidn/NEOMOD\_Simulator} and GitHub.}
-- an easy to operate code that can be used to generate user-defined NEO samples from the model. 

The original model, hereafter NEOMOD v1.0 or NEOMOD1 for short, was calibrated on the Mt. Lemmon 
(IAU code G96) and Catalina (703) telescope observations during the 8-year long period from 2005 to 2012.
This was done for two reasons: (1) the photometric sensitivity of G96 and 703 from 2005--2012
was thoroughly characterized in Jedicke et al. (2016), and (2) Granvik et al. (2018) used the same 
dataset to calibrate their NEO model. We improved the methodology and applied it to the same dataset, 
without the need for an extensive work on characterizing the photometric bias. The differences 
between NEOMOD1 and Granvik et al. (2018) therefore entirely reflected the changes in methodology 
(and not observational constraints). The improvements included: (i) cubic splines to represent the 
magnitude distribution of NEOs, (ii) rigorous model selection with \texttt{MultiNest}, (iii) a physical  
model for disruption of NEOs at low perihelion distances (Granvik et al. 2016), (iv) an accurate estimate 
of the impact fluxes on the terrestrial planets, and (v) a flexible setup that can be readily adapted 
to any current or future NEO survey.\footnote{NEOMOD calibration on the ATLAS (Heinze et al. 2021) 
and WISE (Mainzer et al. 2019) observations is under development.}

We found that the sampling of main-belt sources by NEOs is {\it size-dependent} with the $\nu_6$ and 
3:1 resonances contributing $\simeq 30$\% of NEOs with $H = 15$, and $\simeq 80$\% of NEOs with 
$H = 25$. This trend most likely arises from how the small and large main-belt asteroids reach 
the source regions (Paper I). The size-dependent sampling suggests that small terrestrial impactors 
preferentially arrive from the $\nu_6$ source, whereas the large impactors can commonly come from
the middle/outer belt (Nesvorn\'y et al. 2021). The NEOMOD1-inferred contribution of the 3:1 source 
to large NEOs ($H \lesssim 18$) implies that main-belt asteroids should drift toward the 3:1 
resonance at the maximum Yarkovsky drift rates ($\simeq 2 \times 10^{-4}$ au Myr$^{-1}$ for a 
$\simeq 1$-km diameter body at 2.5 au). In Paper I, we therefore suggested that the main-belt asteroids 
on the sunward side of 
the 3:1 resonance ($a<2.5$ au) have obliquities $\theta \simeq 0^\circ$; the ones with $a>2.5$ au 
should have $\theta \simeq 180^\circ$ (in the immediate neighborhood of the resonance). 
These predictions were confirmed from lightcurve observations (\v{D}urech \& Hanu\v{s} 2023).
We verified the size-dependent disruption of NEOs at small perihelion distances (Granvik et al. 
2016), and found a similar dependence of the disruption distance on the absolute magnitude.

Here we extend NEOMOD to incorporate new data from the G96 telescope (hereafter NEOMOD v2.0 or 
NEOMOD2 for short). The camera of G96 was upgraded to a wider field of view (FoV; $2.23^\circ 
\times 2.23^\circ$) in May 2016 and the G96 telescope detected 11,934 unique NEOs between May 31, 
2016 and June 29, 2022 (Fig. \ref{detections}). This can be compared to only 2,987 
unique NEO detections of G96 for 2005--2012 ($1.1^\circ \times 1.1^\circ$ FoVs). 
For completeness, we also include 3,057 unique NEO 
detections of G96 between January 2, 2013 and May 16, 2016. The two new observational 
datasets are referred to as the ``new CSS'', whereas the previous dataset used in Paper I is the 
``old CSS''. We do not attempt to combine the old and new CSS datasets in this work, because 
here we develop a new method for characterizing the photometric bias of new CSS (Sect. 2), and 
we do not want to mix the old and new approaches. The detection statistics of new CSS is large 
enough for the new CSS to stand on its own. The 703 telescope did not detect a comparatively 
large number of unique NEOs since 2013 and is not included here.

This article is structured as follows. In Sect. 2, we describe how the photometric bias was 
characterized for the new CSS. Section 3 briefly reviews the methodology that
was borrowed from Paper I, including the definition of NEO sources, $N$-body integrations,
choices of model parameters, and model optimization with {\tt MultiNest}. The final model, NEOMOD2, 
synthesizes our current knowledge of the orbital and absolute magnitude distribution of NEOs 
(Sect. 4). We demonstrate that the population of very small NEOs detected by G96 shows an excess for 
low-eccentricity orbits with $a \simeq 1$--1.6 au and suggest that the excess can be explained if
large NEOs tidally disrupt during close encounters to the terrestrial planets (Sect. 5). Planetary
impacts are discussed in Sect. 6.

\section{Characterizing the observational bias of new CSS}

The G96 telescope has a carefully recorded pointing history, amounting to over 240,000 frames for the 
2013-2022 period. Here we use new detections and incidental redetections of 
NEOs by CSS. We count each individual NEO only once (i.e., as detected) and do not consider 
multiple (incidental or not) detections of the same object. With this setup, we mainly care about 
the detection \textit{probability} of an object by CSS, and not about the number of images in which 
that same object was detected (cf. Granvik et al. 2018). The detection probability 
(or bias for short) of a moving object is defined as the probability that the CSS detection pipeline picks up 
an object in at least three images with the same pointing direction taken by CSS in short succession 
on a single night (image set or frame). The three (or more) tracklets must be correctly linked to count as 
true detection. The detection probability can be split into three parts: (i) the geometric probability 
of the object to be located in the image set, (ii) the photometric efficiency of detecting the 
NEO's tracklet, and (iii) the trailing loss. To account for (i), we use the publicly available 
\texttt{objectsInField}\footnote{\url{https://github.com/AsteroidSurveySimulator/objectsInField}}
code ({\texttt{oIF}) from the Asteroid Survey Simulator (AstSim) package (Naidu et al. 2017).
See the GitHub documentation of oIF for a detailed description of the code and Paper I for the 
implementation used here for NEO modeling.

As for (ii), our starting point is a 3.5 GByte tarball of nearly 10 years of data from the G96 telescope.
To make this tarball, each G96 field was calibrated against Gaia-DR2 stars (Gaia Collaboration et al. 2018) 
and the moving object identification was done against the most recent MPCORB catalog (as of October 2022). 
For each frame, a list is provided of both the objects that were identified in the G96 image, and those that were expected 
to be in the field of view but were not detected. The data start on January 2, 2013 and end on 
June 29, 2022. There is one file per set of images with the same pointing direction 
taken by G96 in a short succession on the same night. The header of each file reports: 
the (1) exposure (typically 30 sec, occasionally 45 sec), (2) number of images in the frame 
(3 to 5, typically 4), (3) MJD when each image was taken, (4) right ascension and declination of the 
image center, (5) image orientation relative to north (always $<$1 deg), and (6) 50\% magnitude value 
($V_{\rm 50}$). The 50\% magnitude value is a variation on a zero-point magnitude calculation.   
We collected all Gaia-DR2 stars that can be identified and scaled them against the matching 
point source SNR values converted to $\Delta$mag. The 50\% value is the magnitude where the 
Gaia-DR2 stars cross an SNR that has been calibrated to give us roughly a 50\% main belt 
asteroid detection rate as determined from a test set used at the time. It is a good 
reference to understand the quality of observing conditions for each frame.

Each file lists all {\it known} main-belt and near-Earth asteroids -- obtained from the MPC catalog 
from October 2022 -- that would appear in G96's frame that night, and specifies whether they were 
detected by the CSS pipeline. The following information is given for each object: the (1) proper motion 
($w$), (2) visual and absolute magnitudes ($V$ and $H$), and (3) semimajor axis ($a$), eccentricity
($e$), and inclination ($i$). The $V$ magnitude was computed from the observing geometry and $H$ 
magnitude reported in the MPC catalog. We discarded nights where fewer than 250 asteroids were
available (detected or not) in all frames taken on the same night of observations (18,509 files in total),
because such a small number of objects did not allow us to accurately derive the detection efficiency for 
that night (Sect 2.1). We also excluded 27 files with fewer than three images per frame (three 
images are required for detection). This left us with 223,865 files in total (one file for each exposure),
61,585 for nights before May 16, 2016 (hereafter CSS1) and 162,280 after May 31, 2016 (hereafter CSS2).

The unique NEO detections were extracted from all files. If the same object was detected more 
than once, we only considered the first instance. NEOs detected on discarded nights were
neglected. We also excluded NEO detections with $w>10$ deg/day 
because we were not able to determine the trailing loss for these excessively large apparent 
motions (Sect. 2.2). This left us with 2,619 unique NEOs for CSS1 and 11,471 unique NEOs 
for CSS2 (objects detected by both CSS1 and CSS2 are listed twice, once in each dataset; Fig.
\ref{detections}). The two datasets report the semimajor axis, eccentricity, inclination, and 
absolute magnitude of the NEOs (at the time of detection). Note that the $H$ magnitudes of all objects were
obtained from the 2022 MPC catalog (downloaded on October 19, 2022); this defines the absolute magnitude system used in this 
is work.\footnote{The absolute magnitudes of the detected NEOs were given to two decimal
digits but the second decimal digit was often zero. This happened because the legacy MPC catalogs,
from which some data were imported, listed only one decimal digit. As this would create 
round-off problems with binning, we randomly added $-0.001$ or $+0.001$ to the reported 
magnitudes. This resolves the problem.} Given the previously identified offset of MPC magnitudes 
(Pravec et al. 2012), the 2022 MPC system may still include systematic errors. In addition, 
as the absolute magnitudes of individual objects are updated with each new release of the 
MPC catalog, one has to be careful when comparing the NEOMOD2 results with new MPC releases. 
 
\subsection{Photometric probability of detection}

We adopt the following method to characterize the photometric probability of detection. 
For each asteroid reported in each file, we first compute the visual magnitude offset 
$V'=V-V_{\rm 3rd}$, where $V_{\rm 3rd}$ is the third faintest 50\% magnitude value listed in the 
frame file header ($V_{\rm 50}$).\footnote{We use the usual Pogson's relation to compute 
the visual magnitude of each object. Nominally, we set the slope parameter $G=0.15$ (Bowell 
et al. 1989) but also tested $G=0.24$ (Pravec et al. 2012). The results described 
in Sect. 4 are practically independent of this choice.}   
The reason for the 3rd faintest is that it takes at least three hits for 
a detection. The 3rd faintest field therefore has the main impact on the efficiency. All 
asteroids appearing in the same frame, detected or not, are binned as a function of 
$V'$, defining $N_{\rm all}(V')$. We also bin the number of asteroids {\it detected} by the 
G96 pipeline and denote it by $N_{\rm det}(V')$. The photometric probability of detection 
in a $V'$ bin is simply the ratio $\epsilon(V')=N_{\rm det}(V')/N_{\rm all}(V')$. 
We use 30 bins between $V'_{\rm min}=-6$ and $V'_{\rm max}=1.5$ (i.e., bin size 0.25 mag). This is 
where practically all NEOs were detected by G96. We therefore do not need to characterize the detection 
efficiency for $V'<V'_{\rm min}$ and $V'>V'_{\rm max}$.\footnote{There were some exceptions 
such as (433) Eros; objects detected with $V'>V'_{\rm min}$ or $V'<V'_{\rm max}$ were discarded.} 

In the next step, we need to find a suitable analytic expression that provides a sufficiently
good approximation for $\epsilon(V')$. This is a matter of compromise. On one hand, 
there is a preference for a simple and robust approximation that will always provide a reasonable 
approximation of binned $\epsilon(V')$, even if the statistics on a given night is relatively poor. 
On the other hand, the analytic function must be sufficiently accurate in the whole range 
$V'_{\rm min}<V'<V'_{\rm max}$, including the transition where $\epsilon(V')$ drops near 
the detection limit, such that no artifacts are introduced. The original functional form
that was adopted in Paper I from Jedicke et al. (2016) was
\begin{equation}
\epsilon(V) =  \frac{\epsilon_0}
{ 1 + \exp \left ( {  V - V_{\rm lim}  \over V_{\rm wid} } \right )   }  \ . \label{eq0} 
\end{equation}
After extensive testing, we adopted 
\begin{equation}
\epsilon(V') = \epsilon_0\ \frac{1 - \left ( { V' - V'_0 \over q_{\rm V} }  \right )^2}
{ 1 + \exp \left ( {  V' - V'_{\rm lim}  \over V_{\rm wid} } \right )^{\!\alpha} }  \ . 
\label{photo}
\end{equation}

We now use $V'=V-V_{\rm 3rd}$ instead of $V$ in Eq. (\ref{photo}}).
There are six parameters: $\epsilon_0$, $V'_0$, $q_{\rm V}$, $V'_{\rm lim}$, $V_{\rm wid}$ and $\alpha$. 
The $\alpha$ parameter improves the analytic fit for $V' \rightarrow V'_{\rm max}$, where 
$\epsilon(V')$ rapidly drops toward zero. The `squeezed' exponential with $\alpha>1$
for $V' > V'_{\rm lim}$ matches this fall off better than a normal exponential. This behavior 
cannot be mimicked by adopting a smaller value of $V_{\rm wid}$ because this would damage the fit 
for $V' < V'_{\rm lim}$ (we fix $\alpha=1$ for $V' < V'_{\rm lim}$).
The quadratic term in the numerator of Eq. (\ref{photo}) was taken 
from Tricarico et al. (2016). It improves the behavior of the analytic fit for $V'_0 < V' < V'_{\rm lim}$, where 
$\epsilon(V')$ has a bending profile that differs from an exact exponential. When the 
statistics on a given night is relatively small (i.e., low $N_{\rm all}$), the bins near 
$V'_{\rm min}$ are sparsely filled, and this would adversely affect the quadratic term, if the
fit is given this much freedom. For $V' < V'_0$, where $V'_{\rm min}<V'_0<V'_{\rm lim}$ is a 
free parameter of the fit, we therefore set $\epsilon(V') = \epsilon_0$. 

The optimization of six photometric parameters was performed with the Simplex method (Press et 
al. 1992). We have the capability to execute the fit for one frame, for all CSS observations 
(Fig. \ref{css}), and everything in between. We find that the number of asteroids in a single 
frame is typically too small for a robust determination of $\epsilon(V')$ in each frame. We therefore 
need to group frames together. Grouping too many frames together would not be optimal, however, 
because the atmospheric conditions may significantly vary between different nights, the observing strategy 
and parameters change over time, etc. We thus choose to characterize $\epsilon(V')$ on a nightly 
basis (see Fig. \ref{night} for an example).

All frames taken on a single night were collected, the offset $V'=V-V_{\rm 3rd}$ was applied 
individually for each frame, but the binning and Simplex fit were done only once for the whole night.
We discarded nights with $N_{\rm all}<250$ because we did not have confidence in the 
results when the total asteroid sample on that night was small. The six photometric parameters 
were individually obtained for 602 nights of CSS1 and 1110 nights of CSS2 (Fig.~\ref{photop}).
Table~1 lists the global photometric parameters for CSS1 and CSS2 for reference. In general, CSS2 
has brighter values of $V'_{\rm lim}$ than CSS1. This means that, for CSS2, $V_{\rm 3rd}$ is a better 
proxy for where the photometric detection efficiency drops. The $V_{\rm 3rd}$ values of CSS2 are 
generally fainter, by a fraction of magnitude, than those of CSS1.
We stress that, even if the six photometric parameters $\epsilon_0$, $V'_0$, $q_{\rm V}$, 
$V'_{\rm lim}$, $V_{\rm wid}$ and $\alpha$ have fixed values for a given night, $V_{\rm 3rd}$ is 
treated individually for each frame. We therefore have an approximate characterization of 
the photometric efficiency on a frame-to-frame basis.\footnote{Note that the method described 
here accounts for the reduction of the detection probability from the camera's {\it fill} factor --
the fraction of the FoV where camera is actually sensitive (parts of the camera are not sensitive
because of gaps, masked pixels, etc.). The the fill factor is implicitly accounted for as 
the detection probability is inferred from detections and non-detections of {\it real} 
objects appearing in each image.}

\subsection{Trailing loss}

The trailing loss stands for a host of effects related to the difficulty of detecting fast 
moving objects. If the apparent motion is high, the object's image (a streak) is smeared over 
many CCD pixels, which diminishes the maximum brightness and decreases S/N. Long trails may be 
missed by the survey's pipeline (due to streaking), the object may not be detected in three 
images of the same frame (as required for a detection), or the streaks in different images may 
not be linked together. The trailing loss is especially important for small NEOs, which can 
only be detected when they become bright, and this typically happens when they are moving very 
fast relative to Earth during a close encounter. 

It is not easy to accurately characterize the trailing loss from the CSS data that are available 
to us. This is mainly because the number of detected objects in CSS frames rapidly falls 
off for high rates of motion. The statistics therefore becomes progressively worse as we 
consider higher and higher rates of motion. Ideally, we would like to investigate different 
effects (see above) separately, because some should vary with cadence, while the trailing 
loss itself (smearing) depends on the angular velocity. This is unfortunately not possible 
because there is simply not enough data for high rates of motion. In addition, we would like 
to characterize the trailing loss on a nightly basis, on a monthly basis, or at least separating 
CSS1 and CSS2. This is also not possible because there is not a sufficient number of 
detections in CSS1 for $w>3$ deg/day.  

We therefore adopt the following (approximate) procedure. We first clump all the CSS1 and CSS2 
observations together and separate $N_{\rm all}$ and $N_{\rm det}$ into 1 deg/day bins in the 
apparent motion, from $w=0$ to $w=10$ deg/day. There are only under 200 unique detections in individual 
bins for $w>10$ deg/day, and that is clearly not good enough for characterizing the trailing loss. 
The asteroids detected with $w>10$ deg/day were discarded from the detection probability computation 
and from the list of detected NEOs. We only consider $w<10$ deg/day. With the 
$w$ binning, the detection efficiency is now $\epsilon(V',w)$. As before (Sect. 2.1), we use 
the Simplex method and Eq. (\ref{photo}) to analytically 
parameterize $\epsilon(V',w)$, and derive the six photometric parameters, which are now 
global for the new CSS, but depend on $w$. The photometric parameters were plotted as a 
function of $w$ to give us sense of how they change and what analytic functions would capture 
that behavior.

An example for $V'_{\rm lim}(w)$ is shown in Fig. \ref{trail}. Even though the dependence of $V'_{\rm lim}$
on $w$ is uneven, we find that $V'_{\rm lim}(w)$ slightly increases to $w \simeq 3.5$ deg/day and then 
drops for $w > 3.5$ deg/day. This means that the detection efficiency improves for the apparent motions approaching 
$w \simeq 3.5$ deg/day, which corresponds to $\simeq 3$ pixels/exposure for CSS2. Confusion of moving 
objects with faint stars probably decreases the detection rates for very slow apparent 
motions.\footnote{We looked into this in more detail and found that the CSS2 detection probability
drops for $w < 0.12$ deg/day. This happens because an object moving this slow appears in only a few pixels 
of the image set, and this greatly diminishes its detection probability. We therefore used $w > 0.12$ 
deg/day for the computation of $\epsilon(V')$ in Sect. 2.1. According to our tests, however, including 
$w < 0.12$ deg/day would not have a significant impact on the overall results described in Sect. 4.}
As we go faster than 3 pixels/exposure there is a double penalty of losses from trailing and the increased 
angular distance between the first and last point (which makes it more difficult to uniquely link 
the observations to a moving object). That could explain why $V'_{\rm lim}(w)$ slopes downward 
for $w > 3.5$ deg/day.

We analytically approximate $V'_{\rm lim}(w)$ as
\begin{equation}
V'_{\rm lim}(w)=V'_{\rm lim}(0) + A w
\end{equation}
for $w<w_1$ and 
\begin{equation}
V'_{\rm lim}(w)=V'_{\rm lim}(0) + A w_1 + 2.5 \log_{10}[1+C(w-w_1)]
\label{loss}
\end{equation}
for $w_1<w<10$ deg/day, and find $A=0.052$, $C=0.192$, $w_1=3.6$ deg/day. To respect the photometric 
conditions of each night, we set $V'_{\rm lim}(0)$ to be equal to $V'_{\rm lim}$ derived for that
night (top-left panel of Fig. \ref{photop}). 

The functional form of trailing loss in Eq. (\ref{loss}) was obtained from the following reasoning.
Let $\phi$ be the characteristic angular dimension of the point spread function (PSF). Let the 
angular rate of motion of the object be $w$ during an exposure time $t$. Assume that trailing 
effects only become important after an object has moved through an angle $\theta=w_1 t$ 
where $w_1$ is identified as the minimum rate of motion at which trailing loss becomes apparent.
Let the flux within the PSF from a stationary source be $f_s=1$. If the same source is moving at 
a rate $w$ for a time $t$ across the image plane, its flux will be spread along a trail of 
angular length $\ell=\phi + w t$. Then the flux within a PSF area along the trail is roughly 
\begin{equation}
    f_t = \frac{ \phi + ( w - w_1 ) t }{ \phi } = 1 + \frac{t}{\phi} ( w - w_1 ).
\end{equation}
Thus, the change in apparent magnitude in a PSF-like region due to trailing is given by
\begin{equation}
\Delta V = 2.5 \log_{10}\bigg[ 1 + \frac{t}{\phi} ( w - w_1 ) \bigg].
\end{equation}
A good rule-of-thumb is that $w_1$ is the rate at which an object moves a full PSF during 
the exposure time.  The G96 PSF is roughly $3\arcsec$ and $t=30\sec$, so we expect $w_1 \sim 
2.4$ deg/day and $\frac{t}{\phi} \sim 0.42$ day/deg, in rough agreement with the 
fitted values (see above). 

In an actual survey system there are many different, often competing, factors at play in the detection 
efficiency including, but not limited to, the ability of the system's software to detect sources in 
an image as a function of the source's shape and an object's rate of motion.  Distant objects, or 
even nearby objects at their stationary points, may move too slowly to be detected as moving between 
successive images.  Sources that trail just a little might be easier to detect than sources that 
trail a little less.  These effects are difficult to calculate from theory so we generalize the 
trailing loss function in Eq. \ref{loss} and fit for the parameters $A$, $C$ and $w_1$.

A similar analysis was performed for other photometric parameters as well. We found that $V_{\rm wid}(w)$ 
can be adequately approximated by $V_{\rm wid}(0)$ for $w<w_2$ with $w_2 \simeq 7$ deg/day. For 
$w>w_2$, the transition from high to low detection detection probabilities near $V'_{\rm lim}(w)$
becomes a step-like function; we thus have $V_{\rm wid} = 0$ for $w>7$ deg/day. The last issue arises 
as there were no objects detected for $V'$ exceeding a certain limit, $V'_{\rm cut}$,
where $V'_{\rm cut} = 1.5$ mag for $w \simeq 0$ (our usual cutoff) and $V'_{\rm cut} = 0$ when 
$w$ approaches 10 deg/day. In the final algorithm for the trailing loss, we implemented this cutoff 
by setting $\epsilon(V')=0$ for $V'>V'_{\rm cut}$.       

\subsection{Detection probability as a function of $a$, $e$, $i$ and $H$}

The detection probability of new CSS, ${\cal P}(a,e,i,H)$, needs to be computed as a function of
$a$, $e$, $i$ and $H$. As we described in Paper I, the model distribution of NEO orbits is 
binned (we use the same binning as in Paper I). We therefore need
to compute ${\cal P}(a,e,i,H)$ in each bin. For each bin, we generated a large number ($N_{\rm obj}=10,000$; 
the required number was determined by convergence tests) of test objects with a uniformly random 
distribution of $a$, $e$ and $i$ within the bin boundaries. The mean anomaly, argument of perihelion, 
and longitude of ascending node 
were randomly chosen between 0 and 360$^\circ$. The \texttt{oIF} code (Naidu et al. 2017) was then 
used to determine the geometric detection probability in each frame. For each $H$ bin, we assigned the corresponding 
absolute magnitude to 10,000 test NEOs and propagated the information to compute the detection efficiency 
$\epsilon_{j,k}(V,w)$, individually for every bin $j$ and frame $k$ (Eq. \ref{photo} and Sect. 2.2). See 
Sect. 4.5 in Paper I for more details.
  
The detection probability ${\cal P}(a,e,i,H)$ is defined as the mean detection probability of 
an object with $(a,e,i,H)$ over the whole duration of each survey. We compute the mean detection probability 
as 
\begin{equation}
{\cal P} = \frac{1}{N_{\rm obj}}\sum_{j=1}^{N_{\rm obj}} \bigg\{1-\prod_{k=1}^{N_{\rm frame}} [1 - \epsilon_{j,k}]\bigg\}\ ,
\label{calp}
\end{equation}
where $N_{\rm frame}$ is the number of frames, and the product of $1 - \epsilon_{j,k}$ over frames stands for 
the probability of {\it non}-detection of the object $j$ in the survey. We compute ${\cal P}$ separately for 
CSS1 and CSS2.

Figures \ref{bias2} and \ref{bias3} illustrate the CSS bias. The detection probability of CSS2 is $\gtrsim 
0.7$ for large, $H \simeq 15$ NEOs, except for those on orbits with $a < 0.8$ au. Fainter NEOs are detected 
with lower probability. Interestingly, ${\cal P}$ shows dips and bumps as a function of NEO's semimajor axis 
(Fig. \ref{bias3}). The dips, where the detection probability is lower, correspond to the orbital periods that
are integer multiplies of 1 year. This is where the synodic motion of NEOs allow them to hide and often not
appear in the survey's frames. This effect has been reported before (Tricarico 2017 and Paper I). 

\section{NEO model parameters and optimization}

The source populations and integration method used to generate the orbital distribution of NEOs from 
each source were described in Paper I. We have 12 sources in total: eight individual resonances ($\nu_6$, 3:1, 
5:2, 7:3, 8:3, 9:4, 11:5 and 2:1), weak resonances in the inner belt, two high-inclination sources (Hungarias 
and Phocaeas), and comets. The integration output was used to define the binned orbital distribution of 
NEOs from each source $j$, ${\rm d} p_j(a,e,i) = p_j(a,e,i)\ {\rm d}a\, {\rm d}e\, {\rm d}i$, and 
normalized it to one NEO,  
\begin{equation}
\int_{a,e,i} p_j(a,e,i)\ {\rm d}a\, {\rm d}e\, {\rm d}i = 1\ ,      
\label{one}
\end{equation}   
effectively representing the binned orbital PDF (probability density function). We used the 
orbital range $a<4.2$ au, $q<1.3$~au, $e<1$ and $i<90^\circ$, hereafter the NEO model domain. This is 
where practically all NEOs detected by new CSS reside.\footnote{Exceptions are: (343158) Marsyas 
with a retrograde orbit and $a=2.527$ au, (3552) Don Quixote, 2019 PR2, 2019 QR6 and three other 
(weakly active) comets on Jupiter-crossing orbits with $a>4.2$ au.} As the binning is done only in $a$, 
$e$, and $i$, the model ignores any possible correlations with the orbital angles (nodal, perihelion 
and mean longitudes). There are 42 bins in $a$, 20 bins in $e$ and 22 bins in $i$,
and 52 bins in $H$ for $14<H<28$.  

We use {\tt MultiNest} to perform the model selection, parameter estimation and error analysis 
(Feroz \& Hobson 2008, Feroz et al. 2009).\footnote{\url{https://github.com/farhanferoz/MultiNest}} 
{\tt MultiNest} is a multi-modal nested sampling routine (Skilling et al. 2004) designed to compute 
the Bayesian evidence in a complex parameter space in an efficient manner. The log-likelihood in 
{\tt MultiNest} is defined as 
\begin{equation}
{\cal L} = \ln P = - \sum_j \lambda_j + \sum_j n_j \ln \lambda_j \ ,
\label{like}
\end{equation}
where $n_j$ is the number of objects detected by CSS in the bin $j$, $\lambda_j$ is the number of objects 
in the bin $j$ expected from the biased model, and the sum is executed over all bins in $a$, $e$, $i$ and $H$.
This definition is identical to that used in Paper I. For two or more surveys, $\cal{L}$ is simply the 
sum of individual survey's log-likelihoods. As we treat CSS1 and CSS2 as two independent surveys,
we have $\cal{L}=\cal{L}_{\rm CSS1}+\cal{L}_{\rm CSS2}$.  

There are three sets of priors: (1) coefficients $\alpha_j$ that determine the strength of different
sources, (2) parameters related to the absolute magnitude distribution, and (3) priors that define 
the disruption model (Granvik et al. 2016).

As for (1), the intrinsic orbital distribution of model NEOs is obtained by combining $n_{\rm s}$ 
sources: $p(a,e,i)=\sum_{j=1}^{n_{\rm s}} \alpha_j\, p_j(a,e,i)$ with $\sum_{j=1}^{n_{\rm s}} \alpha_j = 1$. 
The coefficients $\alpha_j$ represent the relative contribution of each source to the NEO population
(i.e., the fraction of NEOs from the source $j$). As the contribution of different sources to NEOs 
may be size dependent (Paper I), we set $\alpha_j$ coefficients to be functions of the absolute magnitude. 
For simplicity, we adopt a linear relationship, $\alpha_j=\alpha_j^{(0)} + \alpha_j^{(1)}(H - H_\alpha)$, 
where $H_\alpha$ is some reference magnitude, and $\alpha_j^{(0)}$ and $\alpha_j^{(1)}$ are new model 
parameters. In practice, we set $\alpha_j(H_{\rm min})$ and $\alpha_j(H_{\rm max})$ for some minimum 
and maximum absolute magnitudes (e.g., $H_{\rm min}=15$ and $H_{\rm max}=28$), and linearly interpolate 
between them. This automatically assures that $\sum_j \alpha_j(H)=1$ for any $H_{\rm min} \leq H \leq H_{\rm max}$.      

As for (2), the differential and cumulative absolute magnitude distributions are denoted by 
${\rm d}n(H) = n(H) {\rm d} H$ and $N(H)$, respectively. The differential magnitude distribution 
produced by source $j$ is set to be ${\rm d}n_j(H) = \alpha_j(H) n(H) {\rm d} H$. The magnitude 
distributions of different sources are similar, but change with $\alpha_j(H)$, which are assumed 
to linearly vary with $H$ (see above). When the contribution of different sources is combined, we 
find that $\sum \alpha_j(H) n(H) {\rm d} H = n(H) {\rm d} H$, which means that $n(H)$ stands for 
the absolute magnitude distribution of the whole NEO population. 

We use cubic splines to represent $\log_{10} N(H)$ (Paper I). The magnitude interval of interest, 
$15<H<28$, is divided into six segments. There are six parameters defining the average slope 
in each segment, $\gamma_j$, and one parameter that provides the overall calibration. We use 
$N_{\rm ref}=N(H_{\rm ref})$ with $H_{\rm ref}=17.75$ (diameter $D=1$ km for the reference albedo $p_{\rm V}=0.14$). 
The normalization constant and slope parameters are used to compute $\log_{10} N(H)$ at the boundaries 
between segments; cubic splines are constructed from that (Press et al. 1992). The splines assure 
that $N(H)$ smoothly varies with $H$. The known sample of NEOs with $H<15$ is thought to be (nearly) 
complete, and there were $\simeq50$ such objects in the MPC catalog from October 2022. We therefore fix 
$N(15)=50$ and compute the $\gamma_1$ slope such that this additional constraint is satisfied.

As for (3), following Granvik et al. (2016), we eliminate test bodies when they reach the critical distance 
$q^*$ ($q^*$ is the perihelion distance below which NEOs completely disintegrate in catastrophic
breakups). Here we assume that the $q^*$ dependence on $H$ is (roughly) linear, and parameterize it by 
$q^* = q_0^* + \delta q^* (H-H_q)$, where $H_q=20$. We use uniform priors for the two parameters, 
$q_0^*$ and $\delta q^*$. To construct the orbital distribution for any $q^*<0.4$~au, we first produce 
the binned distributions (from each source) for $q^*=0$, 0.05, 0.1, 0.15, 0.2, 0.25, 0.3, 0.35 and 0.4 au. 
The fitting routine then linearly interpolates between these distributions to any intermediate value 
of $q^*(H)$. The resulting orbital distribution, $p_{q^*}$, which now also depends on the absolute 
magnitude, $p_{q^*}=p_{q^*}(a,e,i,H)$, is normalized to 1 ($\int p_{q^*}(a,e,i,H)\ {\rm d}a\, {\rm d}e\, 
{\rm d}i = 1$ for any $H$). 
         
In summary, our biased NEO model is
\begin{equation}
{\cal M}_{\rm b}(a,e,i,H) =  n(H)\, {\cal P}(a,e,i,H) \sum_{j=1}^{n_s} \alpha_j(H)\, p_{q^*,j}(a,e,i,H)\, \ , 
\label{model}
\end{equation}
where $\alpha_j$ are the magnitude-dependent weights of different sources ($\sum_j \alpha_j(H) = 1$), 
$n_s$ is the number of sources, $p_{q^*,j}(a,e,i,H)$ is the PDF of the orbital distribution of NEOs 
from the source $j$, including the size-dependent disruption at the perihelion distance $q^*(H)$ (this
is the only $H$-dependence in the $p$ functions), 
$n(H)$ is the differential absolute-magnitude distribution of the NEO population (the log-cumulative
distribution is given by splines), and ${\cal P}(a,e,i,H)$ is the CSS detection probability
(Eq. \ref{calp}). For each \texttt{MultiNest} trial, Eq.~(\ref{model}) is constructed by the methods
described above. This defines the expected number of events $\lambda_j = {\cal M}_{\rm b}(a,e,i,H)$ 
in every bin of the model domain, and allows \texttt{MultiNest} to evaluate the log-likelihood 
from Eq. (\ref{like}).

The intrinsic (debiased) NEO model is simply
\begin{equation}
{\cal M}(a,e,i,H) =  n(H)\, \sum_{j=1}^{n_s} \alpha_j(H)\, p_{q^*,j}(a,e,i,H)\, \ . 
\label{model2}
\end{equation}  
By integrating Eq. (\ref{model2}) over the orbital domain, given that 
$\int p_{q^*,j}(a,e,i,H) \ {\rm d}a\, {\rm d}e\, {\rm d}i = 1$ and $\sum_j \alpha_j(H) = 1$, 
we verify that $n(H)$ stands for the (differential) magnitude distribution of the whole NEO population. 

\section{NEOMOD v2.0}

Our base NEO model accounts for $n_{\rm s}=12$ sources (Paper I). Each source has a magnitude-dependent 
contribution (Sect. 3) and the source weights $\alpha_j(15)$ (for $H=15$) and $\alpha_j(28)$ (for $H=28$)
therefore represent $2(n_{\rm s}-1)$ model parameters (the last source's contribution is computed from
$\sum_{j=1}^{n_{\rm s}} \alpha_j = 1$). There are six parameters related to the magnitude distribution,
$N_{\rm ref}$ and $\gamma_j$, $2 \leq j \leq 6$ ($15 \leq H \leq 28$).\footnote{We tested different 
sectioning of the magnitude range and found that having six intervals $H=15$--16.5, 16.5--17.5,
17.5--20.0, 20.0--24.0, 24.0--25.0, and 25.0--28.0 works slightly better than having equal spacing.}
The $\gamma_1$ parameter is fixed such that $N(15)=50$. In addition, the $q_0^*$ and $\delta q^*$ 
parameters define the disruption model. This adds to 30 model parameters in total. We used uniform 
priors for all parameters (see Paper I for the multivariate uniform distribution of $\alpha_j(15)$ and 
$\alpha_j(28)$). The CSS fits were executed with the \texttt{MultiNest} code (Sect. 3).
The orbital distribution of NEOs from the best-fit (i.e., highest-likelihood) intrinsic model 
${\cal M}$ is shown in Fig. \ref{unb}. 
The NEOMOD Simulator (see Paper I) was updated and is available for 
download.\footnote{\url{https://www.boulder.swri.edu/\~{}davidn/NEOMOD\_Simulator} and GitHub.} 

\texttt{MultiNest} provides the posterior distribution of model parameters. The results are generally 
consistent with those of Paper I, but there are also several interesting differences (Table 2). As before, 
we only have upper bounds on the contribution of 7:3, 9:4 and JFC sources. The models without these sources, 
however, are disfavored at $\Delta \ln {\cal Z} > 9.2$ (Bayes factor). We thus prefer to keep these 
sources in the base model. The $\nu_6$ source now has a lower contribution for $H=15$ ($0.06 \pm 0.03$ 
vs. $0.12 \pm 0.06$ in Paper I) and a higher contribution for $H=28$ ($0.60 \pm 0.02$ vs. $0.42 
\pm 0.04$ in Paper I). The opposite happens for the 3:1 resonance, which now has a $0.28 \pm 0.03$ 
contribution for $H=15$ (previously $0.22 \pm 0.04$) and $0.31 \pm 0.02$ contribution for $H=28$ (previously 
$0.34 \pm 0.03$). The contribution of Hungarias for $H=28$ has an upper limit (0.029; previously $0.06 
\pm 0.03$). These differences are most likely related to how the observations of 703 and G96 telescopes
were combined in Paper I (see Sect. 8 in Paper I and the footnote below).
The uncertainties of all parameters are lower than in NEOMOD1, typically by almost a factor 
of 2. The absolute magnitude and disruption parameters are similar to those reported in Paper I. We 
find $N(17.75)=936 \pm 29$ (Table 3).\footnote{In Paper I, we experimented with two approaches to combining 
the data from the 703 and G96 telescopes. In the first one, inspired by Granvik et al. (2018), 
the detection biases of the two telescopes were combined into a joint survey (see Paper I for details). 
Strictly speaking, this is not ideal because the detection bias of the G96 survey only applies to NEO 
detections in the G96 survey (and not 703), and vice versa. We verified in Paper I that the joint-survey 
approach gives $N(17.75)<1000$ (Granvik et al. (2018) estimated $N(17.75)=962^{52}_{-56}$) even if both 703 
and G96 -- when considered separately -- give $N(17.75)>1000$ (for old CSS and the bias from Jedicke 
et al. (2016)). In the second and more accurate method, 703 and G96 were treated separately in 
\texttt{MultiNest} and were combined at the log-likelihood level. This, however, produced $N(17.75)=1010 
\pm 19$ in Paper I. Here we find that these model inconsistencies most likely reflected a slight 
inaccuracy of the observational bias reported for old CSS in Jedicke et al. (2016).}   
 

The biased best-fit model ${\cal M}_{\rm b}$ is compared to CSS NEO detections in Fig. \ref{bmodel}.
The distributions in Fig. \ref{bmodel} are broadly similar. There seems to be a slight excess of 
CSS NEO detections with $q \sim 1$ au and $1<a<1.6$ au. The 1D PDFs in Figs. \ref{bright} and 
\ref{faint} show the comparison in more detail. For relatively bright NEOs ($15<H<25$; Fig. 
\ref{bright}), ${\cal M}_{\rm b}$ is statistically indistinguishable from CSS detections. The 
Kolmogorov-Smirnov (K-S) test (Press et al. 1992), applied to the four 1D distributions in Fig. \ref{bright},
shows that the null hypothesis (the distributions are drawn from the same underlying distribution) 
cannot be rejected (K-S probability $p>0.05$). The troughs in the semimajor axis distribution at 
$a\simeq1.6$ and 2.1 au are produced by the lower detection efficiency of CSS for orbital periods 
near 2 and 3 years (synodic effect; Fig. \ref{bias3}). The tiny excess of NEOs detected by CSS  with 
$i=20$--30$^\circ$ (red line in Fig. \ref{bright}c) can be related to the contribution of 
high-inclination sources (Hungarias or Phocaeas).  

For faint NEOs ($25<H<28$; Figs. \ref{faint}), ${\cal M}_{\rm b}$ is indistinguishable from 
CSS detections in $i$ and $H$, but there is a major discrepancy in $a$ and $e$, where the CSS detections
show a large excess for $1<a<1.6$ au and $e<0.4$. The 1D K-S tests applied to the $a$ and $e$ distributions 
indicate that the null hypothesis can be rejected ($p<10^{-5}$). The same problem was already discussed in 
Paper I, where we verified that the excess cannot be explained by a rapid drift of $D <100$ m asteroids 
across the $\nu_6$ resonance. The excess also cannot be related to disruption of NEOs at low perihelion
distances (Granvik et al. 2016 and Paper~I), because (i) NEOs with $1<a<1.6$ au and $e<0.4$ do not reach
very low perihelion distances, and (ii) we need to add objects to our model, and not remove them, to 
explain the excess of detections. This problem is most likely related to {\it tidal disruption} of large
NEOs during planetary encounters (Granvik \& Walsh 2017, 2022, 2023); a relatively large fraction ($\simeq 20$--30\%) 
of small NEOs with $25<H<28$, $1<a<1.6$ au and $e<0.4$ can be fragments of tidally disrupted NEOs.
We discuss this issue in Sect. 5.      

The intrinsic (debiased) absolute magnitude distribution from our base model ${\cal M}$ is shown in 
Fig.~\ref{harris}. It is nearly identical to that reported in Harris \& Chodas (2021; hereafter HC21) 
for $H<25$. There is a large difference between ${\cal M}$ and HC21 for $H>25$, where 
the NEOMOD2 distribution has a well defined slope index $\gamma \simeq 0.51$
(equivalent to a power index $\simeq 2.6$ of the cumulative size distribution). Here the distribution given in HC21 
is significantly steeper ($\gamma \simeq 0.62$ for $24<H<27$ or even $\gamma \simeq 0.75$ 
for $H>26$). The same discrepancy was already noted in NEOMOD1 -- here we confirm it from a detailed 
analysis of new CSS. The slope of our size distribution for $H>25$ is consistent with the slope
expected for a population that reached the collisional equilibrium (Dohnanyi 1969). The steeper
slope in HC21 (cumulative size index $\simeq 3.75$ for $H>26$) would require some additional 
explanation.
 
For reference, HC21 obtained $2.44 \times 10^7$ NEOs with $H<27.75$ whereas we only have $0.912 \times 
10^7$ NEOs with $H<27.75$ - a multiplicative factor of $\simeq 2.7$ difference (Table 3). It is
possible that we overestimated the CSS detection efficiency by a factor of $\sim 2$--3 for 
$H \simeq 28$. If so, this would bring our magnitude distribution up by the same factor. We do not 
believe, however, that this is the case. For example, NEOMOD1 -- where the detection efficiency was
obtained for old CSS (2005-2012) from Jedicke et al. (2016) -- produced practically the same result
as we find here from the new analysis of new CSS (2013-2022). It would be strange if two observational 
datasets and two (independent) analyses of the detection efficiency produce the same error. It is also 
possible that the magnitude distribution reported in HC21, who based their estimate on NEO 
redetections and extrapolated it to $H>25$, is too steep for $H>25$.\footnote{Here we compare our results
  with the case from Harris \& Chodas (2021) where NEOs
  with $H>24$ were given the slope $1.0 (V_{\rm lim} - H)$. This is the theoretically
expected slope and the one that better connects to the bolide data (if a fixed
impact probability is adopted, but see Sect. 6). Harris \& Chodas (2021) pointed out that the slope $0.8 (V_{\rm
lim} - H)$ better matches the slope obtained from their redetection method near
$H=24$. This shallower slope for $H>24$ would be in better agreement with our results.}

The redetection method is limited to a magnitude range where the numbers of new 
detections and redetections are statistically large ($17\lesssim H \lesssim 24$; Harris \& D'Abramo 2015). To extrapolate 
the results to fainter magnitudes, HC21 assumed that a survey detects an increasingly smaller fraction 
of the NEO population and estimated -- from the statistics of close encounters of faint NEOs to the Earth 
-- that this fraction was proportional to $10^{-0.8 H}$. The proportionality was further adjusted to 
$10^{-1.0 H}$ for $H>26$ to better fit bolide observations (Brown et al. 2002, 2013). But HC21 implicitly 
assumed, by anchoring the results to the redetection approach at $H \simeq 24$, that the orbital 
distributions of small and large NEOs are the same. We already showed in Paper I that they are not the same
(also see Granvik et al. 2016, 2018). Moreover, as we discuss in Sect. 5, tidal disruption of large NEOs produce small NEOs 
with orbits that have high probabilities of Earth encounters. It may therefore be somewhat problematic 
to infer the general characteristics of the faint NEO population from the encounter statistics alone.   

We confirm the need for the size-dependent disruption of NEOs at small perihelion distances, as originally 
pointed out in Granvik et al. (2016) and Paper I. Clearly, any model where the disruption is not taken 
into account produces a strong excess of low-$q$ (or high-$e$) orbits. The $q^*(H)$ dependence found 
here, $q^* = 0.135 + 0.032\, (H-20)$ with $q^*$ in au, is somewhat steeper -- implying disruption at larger 
perihelion distances for $H>20$ -- than the one inferred in Granvik et al. (2016). Based on this we 
suggest that small NEOs disrupt at slightly larger perihelion distances than found in Granvik et al. 
(2016).
  
\section{A case for tidal disruption}

We find that the largest excess of CSS NEO detections happens for $1<a<1.6$ au, $q \simeq 1$ au and 
$i \lesssim 10^\circ$ (Fig. \ref{excess}). In Paper I we tested whether small main-belt asteroids 
($D < 100$ m) can drift by the Yarkovsky effect over the $\nu_6$ resonance to directly reach 
the NEO orbits with $1<a<1.6$ au and $e<0.4$, and found the orbital distribution of NEOs constructed 
from the simulation with fast drifts was nearly identical to that obtained for the $\nu_6$ resonance 
with the standard approach. This shows that even very small asteroids cannot pass the $\nu_6$ resonance 
and the excess of faint NEO detections for $25<H<28$ must be related to something else. 

Tidal disruption of NEOs is the main suspect (as originally proposed by Granvik \& Walsh 2017, 2022, 2023). 
The orbits with $1<a<1.6$ au, 
$q \simeq 1$ au and $i \lesssim 10^\circ$ have: (i) large probabilities of having close encounters 
with the Earth (e.g., Fig. 5 in Morbidelli \& Gladman 1998), and (ii) low encounter speeds ($v_\infty 
\lesssim 5$ km/s; Fig. 6 in Morbidelli \& Gladman 1998). This is the situation in which tidal disruptions 
are most likely to happen. For example, Richardson et al. (1998) showed that rubble pile bodies 
catastrophically disrupt (`Shoemaker-Levy-9' type of disruption) for $v_\infty \lesssim 5$ km/s and
encounter distances $d \lesssim 2$ $R_{\rm Earth}$, where $R_{\rm Earth}=6371$~km is the Earth radius.
{\it We therefore propose that the excess of small NEOs identified here ($25<H<28$ or $9<D<36$ m for 
the reference albedo $p_{\rm V}=0.14$) is caused by tidal disruption of $D\gtrsim50$ m NEOs.}

A realistic modeling of tidal disruption would require monitoring close planetary encounters of NEOs 
from each source. Unfortunately, we have not recorded any encounters in the $N$-body simulations described 
in Paper I, and we thus cannot conduct a detailed investigation of tidal disruption here. Instead, we 
performed the following test. NEOMOD works well for $H<25$ (Fig. \ref{bright}). We used the base NEOMOD 
model for $H<25$ and multiplied the intrinsic NEO population in each orbital bin by the probability that a body
in the bin would have a close encounter with the Earth.\footnote{We also built models where the close 
encounters with Venus and Mars were included, in addition to Earth encounters. The results of 
these models are very similar to those discussed here for Earth encounters (the Venus-crossing 
NEO population is relatively small and Mars has a relatively low mass). Here we focus on Earth encounters 
because the excess of small NEOs happens along the $q \simeq 1$ au line.} The probability was computed by the \"Opik 
formalism (Bottke et al. 1994). The resulting orbital distribution, which approximates how fragments of 
tidally disrupted NEOs would populate orbital space, was normalized to one and supplied to 
\texttt{MultiNest} as an additional source. Note that this method ignores the orbital evolution of 
fragments in NEO space. It also assumes that the production of small fragments from tidal disruption is a 
steady-state process; this would not be quite right if the contribution of only a few random disruption 
events is important.   

We found that including tidal disruption as an additional source does not change the results
for bright NEOs ($H<25$). This is expected because the model without tidal disruption was able to 
match the orbital and absolute magnitude distribution of bright NEOs (Fig. \ref{bright}), and the 
disruption of a few large asteroids is not expected to significantly change the distribution for $H<25$.
For faint NEOs, however, the best fit requires a significant contribution from tidal disruption. Specifically, 
for $H=28$, \texttt{MultiNest} estimates the tidal disruption weight $\alpha_{\rm td}=0.3 \pm 0.05$.   
The biased best-fit model with tidal disruption is compared to CSS NEO detections in Figs. \ref{tidal}. 
This plot can be contrasted with Fig. \ref{faint} where tidal disruption was ignored. We see that the fit
has substantially improved. The excess for $1<a<1.6$ au and $e<0.4$ has nearly disappeared -- both
the semimajor axis and eccentricity distribution show the overall shapes that match observations
much better than in Fig. \ref{faint}.\footnote{The semimajor axis distribution in Fig. \ref{tidal}a 
can formally be rejected (based on a K-S test), because the biased model distribution is too strongly 
peaked near 1 au, whereas the CSS detections peak near 1.3~au. Some of our test idealizations can 
be responsible for this. For example, we adopted a steady state and ignored the orbital evolution of 
fragments.} This suggests that we are on the right track to resolve this
problem (Granvik \& Walsh 2017, 2022, 2023). The absolute magnitude distributions of NEOs with and without
tidal disruption are practically the same. For example, $\gamma_6=0.53\pm0.01$ with tidal disruption   
and $\gamma_6=0.509\pm0.005$ in the base model without tidal disruption. This means that the 
magnitude distribution difference for $25<H<28$ between HC21 and this work is not resolved when 
the effects of tidal disruption are (approximately) taken into account. A more realistic modeling of 
tidal disruption is left for future work.

Accurate modeling of tidal disruption will need to account for the interior structure
of NEOs. There is evidence that the interior structure changes for NEOs with $D \simeq 100$ m
(roughly $H \simeq 23$). For $D>100$ m, asteroids do not have -- with some
exceptions-- spins faster than $\sim 10$ rotations/day (spin period $\sim
2.5$ hours). This "spin barrier" most likely indicates that $D>100$ m
asteroids do not have large tensile strengh, and are hold together by gravity
(Pravec \& Harris 2000). For $D<100$ m, however, the spins can be as fast as
$\sim 1000$ rotations per day, indicating that these smaller bodies must often have
substantial strength and that their internal structure is probably akin to that
of consolidated rock (monolith). This has important implications for tidal
disruption. Specifically, the weak NEOs with $D>100$ m could be relatively
easily disrupted during close planetary encounters, whereas the stronger NEOs
with $D<100$ m should survive more often. This could help to explain some of
the trends discussed above.

\section{Planetary impacts}

All planetary impacts were recorded by the $N$-body integrator (Paper I). The record accounts for impacts of 
bodies with $q<1.3$ au (NEOs) and $q>1.3$ au (e.g., Mars-crossers). We thus have complete information to 
determine the impact flux on all terrestrial planets, including Mars. We followed $10^5$ test 
bodies from each source and have good statistics to determine the impact flux of NEOs even from 
distant main belt sources (e.g., 9:4, 2:1). To combine impacts from different sources, we compute 
the total impact flux, $F_{\rm imp}$, from  
\begin{equation}
F_{\rm imp} = n(H) \sum_{j=1}^n \alpha_j(H) {p_{{\rm imp},j}(q^*(H)) \over \tau_j (q^*(H))}\ ,  
\label{fimp}
\end{equation}  
where $n(H)$ is the best-fit absolute magnitude distribution of NEOs, $\alpha_j(H)$ are the 
magnitude-dependent source weights (Table 2), $p_{{\rm imp},j}$ is the probability of planetary 
impact for each body inserted in the source $j$, and $\tau_j$ is the mean lifetime of NEOs 
evolving from the source $j$. Parameters $p_{{\rm imp},j}$ and $\tau_j$ depend on $q^*$ and are 
therefore also a function of $H$ (via the linear relationship between $q^*$ and $H$, as defined 
by the best--fit model). We reported them for a reference value $q^*=0.1$ au in Table 5 in Paper I. 
 
Figure \ref{impacts} shows $F_{\rm imp}(H)$, converted to a cumulative distribution, for the terrestrial 
planets. For comparison, we also plot the impact flux on the Earth from HC21 who estimated it by multiplying their absolute 
magnitude distribution $n(H)$ (illustrated in Fig. \ref{harris}) by a constant (i.e., magnitude 
independent) impact probability $P_{\rm i}=1.5 \times 10^{-3}$ Myr$^{-1}$ (Stuart 2001, Harris \& D'Abramo 2015). We 
confirmed in Paper I that this is a correct assumption for large NEOs ($H \lesssim 20$), and only for large NEOs (see below). 
Consistently with Paper I, here we find that the average interval between impacts of $H<17.75$ NEOs ($D>1$ km for 
$p_{\rm V}=0.14$) is 650 kyr. Applying the same fixed impact 
probability to small NEOs, HC21 found that the average interval between impacts of $H<28$ NEOs 
(roughly $D>10$ m for $p_{\rm V}=0.14$) is $\simeq 19$ yr. In Paper I, we already explained 
that the impact probability changes with absolute magnitude; this happens because the $\nu_6$ 
resonance -- known for its high impact probability (Table 5 in Paper I) -- is an important 
source of small NEOs. In the case without tidal disruption, here we find 
$P_{\rm i}=2.9 \times 10^{-3}$ Myr$^{-1}$ for $H=28$ (nearly two times the nominal) and the 
average interval between impacts $\simeq 29$ yr (HC21 population is $\simeq 3$ times higher 
for $H<28$ but the impact probability is $\simeq 2$ times lower). With tidal disruption, 
the average interval between impacts of $H<28$ NEOs is $\simeq 17$ yr. 

An interesting difference between HC21 and this work is identified for intermediate-size NEOs 
($20<H<26$; Fig. \ref{impacts}). For example, HC21 estimated that the mean time between impacts of $H<22$
NEOs ($D>140$ m for the reference albedo $p_{\rm V}=0.14$) is $\simeq 37$,000 yr, whereas we find 
$\simeq 21$,400 yr. This is contributed by two factors: (1) our population of $H<22$ NEOs is slightly larger 
that the one reported in HC21 (Fig. \ref{harris}), and (2) our impact probability for $H<22$ NEOs 
is slightly higher ($P_{\rm i}=2.4 \times 10^{-3}$ Myr$^{-1}$ for $H=22$; due to the larger contribution of the 
$\nu_6$ resonance to small NEOs). Using our estimate and assuming the Poisson statistics, the probability 
of one impact of a $H<22$ NEO on the Earth in the next 1,000 yr is found to be $\simeq 4.5$\%.        

\section{Discussion}
 
\subsection{Terrestrial impacts of small NEOs}

Brown et al. (2002) analyzed satellite records of bolide detonations in the Earth's atmosphere to estimate 
the impact flux of $\sim1$--10 m bodies. For $D \simeq 10$ m, roughly equivalent to $H=28$ for our reference 
albedo $p_{\rm V}=0.14$, the average interval between impacts was found $\simeq 10$ yr (with a factor of 
$\simeq 2$ uncertainty). The infrasound data from Silber et al. (2009), as reported by Brown et al. 
(2013), indicate a somewhat shorter interval but the error bars of these estimates overlap with the bolide 
data. As for fireball events recorded on the CNEOS website,\footnote{https://cneos.jpl.nasa.gov/fireballs/} 
at least three impactors over the past 20 yr,
including the Chelyabinsk meteorite (Brown et al. 2013), had estimated pre-atmospheric-entry diameters 
$D>10$ m. Together, these estimates suggest that the average interval between $D>10$ m impacts is 
$\simeq 10$ yr, or perhaps even somewhat shorter.  

These results motivated HC21 to use a slightly steeper extrapolation of the NEO magnitude distribution to 
$H \sim 28$ such that their impact flux estimate is more in line with impact observations. Here we showed 
that the magnitude distribution is in fact relatively shallow ($\gamma \simeq 0.51$ for $15<H<28$) but
the impact probability on the Earth increases for smaller NEOs (due to preferential sampling of the
$\nu_6$ resonance and tidal disruption). The mean interval between impacts of $H<28$ NEOs is estimated here 
to be $\simeq 17$ yr (Fig. \ref{impacts}). This is a factor of $\gtrsim 1.7$ longer than the estimates 
based on bolides, infrasound and CNEOS. We speculate that the effects of tidal disruption may be even more
important for terrestrial impacts than our simple test in Sect. 5 would indicate. A detailed investigation 
of tidal disruption is left for future work.

\subsection{Lunar/Martian craters}

Our work could explain the difference between the size distributions of lunar and Martian craters (Daubar 
et al. 2022). The recently formed, small Martian craters have relatively shallow size distribution
($\simeq 2.2$ cumulative index from Daubar et al. 2022). For small lunar craters, Neukum et al. (2001) 
reported $\simeq 3.4$ cumulative index for crater diameters $\simeq 0.1$--2 km, which would correspond to 
$\simeq 3$--100 m impactors (the distribution is probably even steeper for smaller impactors; Speyerer
et al. 2020). The size distribution of small lunar craters is thus significantly steeper than the 
size distribution of small Martian craters. Previous work sought to explain this difference by meteoroid 
ablation and fragmentation in Martian atmosphere (Popova et al. 2003), but the atmospheric effect should 
be irrelevant for $D>10$ m impactors. Secondary impacts, which can contribute to the size distribution of 
small craters, should have similar effects for the Moon and Mars. As an important caveat, we note that the 
craters reported in Daubar et al. (2022) are very small, roughly corresponding to $D<10$ m impactors. 

Here we find the cumulative index $\simeq 3.1$ for small lunar impactors and $\simeq 2.5$ for small Martian 
impactors ($25<H<28$ or $9<D<36$ m for $p_{\rm V}=0.14$). The lunar distribution is steeper for two reasons:
(1) Small NEOs preferentially evolve from the $\nu_6$ resonance; this favors lunar impacts because small
objects spend shorter time on Mars-crossing orbits than the large ones (their impact window is short). 
(2) Tidal disruption produces excess of small NEOs ($25<H<28$) for $1<a<1.6$ au, $q \simeq 1$ au and 
$i \lesssim 10^\circ$ (Sect. 5), and these fragments are more likely to hit the Moon (or Earth) than Mars. Tidal 
disruption during close encounters to Mars should happen as well but it is hard to find any evidence 
for that in the CSS data (this may suggest that no large tidal disruption events happened during Mars 
encounters recently).\footnote{Alternatively, small NEOs produced by tidal encounters to Mars do become 
bright enough, given their relatively distant orbits, to be detected by a terrestrial observer.}

For $H \sim 28$, the Mars-to-Moon impact ratio, normalized to the unit surface area, is 
found to be $R_{\rm b}=0.8$ in the model without tidal disruption and $R_{\rm b}=0.5$ in the model 
with tidal disruption. The trend of decreasing $R_{\rm b}$ for smaller impactors is consistent with 
the results reported in Paper I, where we found $R_{\rm b}=1.2$ for $H \sim 25$. For reference, 
Hartmann (2005) and Marchi (2021) adopted $R_{\rm b}=2.6$ for all asteroid impactor sizes when they used 
the lunar chronology for Mars. With our new $R_{\rm b}$ estimates, which imply lower impact flux 
on Mars, the young terrains on Mars dated from $9<D<36$ m impacts should be $\sim 2$--5 times older 
than thought before. 

\subsection{PM excess of meteorite falls}

Tidal disruption could also help to explain the PM excess of meteorite falls (Paper I). The PM/AM ratio measures 
the relative frequency of meteorite falls before (6--12 h) and after (12--18 h) noon. It is usually 
reported as the number of afternoon falls (12--18 h) over the number of day-time falls (6--18 h), to 
express the observed excess of afternoon falls, here denoted as ${\cal E}$. Ordinary chondrites (OCs), 
for example, have ${\cal E}=0.63 \pm 0.02$ (Wisdom 2017), but in Paper I we obtained ${\cal E}=0.47 
\pm 0.02$ and $0.50 \pm 0.05$ for the $\nu_6$ and 3:1 resonances, respectively (also see Morbidelli 
\& Gladman 1998). This difference could be resolved because the excess small NEOs detected by CSS, which 
we attribute to tidal disruption here, happens for $1<a<1.6$ au, $q \simeq 1$ au and $i \lesssim 10^\circ$, 
and impacts from these orbits are expected to have ${\cal E}>0.6$ (Fig. 6 in Morbidelli \& Gladman 
1998). We leave a detailed investigation of this issue for future work.

\subsection{Orbits and CRE ages of meteorites} 

Our work suggests that tidal disruption should be progressively more important for small terrestrial 
impactors and if so, we would expect that many meteorites should have orbits with $1<a<1.6$ au and $q 
\simeq 1$ au. But this does not seem to be the case: the meteorite falls and bolides detected by US 
Government sensors show a broad orbital range with $q=a(1-e)<1$ au and $Q=a(1+e)>1$ au (Brown et al. 2013, 
2016; Granvik \& Brown 2018). It could be that some additional dependencies make a difference. For 
example, the bolide detections should scale with the kinetic energy of the impactor. This means that
disrupted objects, which have relatively low impact speeds and thus lower impact energies, should
have a reduced presence the detected bolide flux. When we fold this 
dependence into the orbital distribution of small impactors, we find that the orbital distribution of
bolides from the model is actually consistent with bolide observations.\footnote{The statistics is not 
ideal because Brown et al. (2016) only reported $\sim 50$ bolide orbits; additional detection biases 
can also be an issue.}   

It needs to be emphasized that the pre-atmospheric-entry diameters of meteorites are characteristically 
$\sim 0.1$ m, which is a factor of $\sim 100$ in size below where we constrained NEOMOD2 from the 
telescopic observations of NEOs ($D\gtrsim 10$ m). The impactor sizes reported from US Government 
sensors are typically $\sim 1$ m (Brown et al. 2016). So, there could also be something problematic 
with extrapolating our expectations from $D\gtrsim 10$ m to $D\lesssim 1$ m.
      
In addition, if many small NEOs were produced in relatively recent tidal disruption events of larger NEOs, 
we would expect that many meteorites would have short Cosmic Ray Exposure (CRE) ages. But the CRE 
distribution of ordinary chondrites does not seem to require any recent tidal disruption events 
(Vokrouhlick\'y \& Farinella 2000; except, perhaps, for the CRE peak of H chondrites at 7--8 Myr). 
It may be the case that tidal disruption affects carbonaceous (C-type) NEOs to a larger degree, because 
they are weaker and have lower density than ordinary chondrites. Noble gas analysis has indicated that 
the CRE ages of most CM and CI chondrites are 0.1--1 Myr, significantly younger than ages of other 
carbonaceous and ordinary chondrites (1--100 Myr; Eugster et al., 2006). Krietsch et al. (2021) found that 
the main CRE age cluster of CMs is at $\simeq 0.2$ Myr and observed further minor peaks at 1, 4.5--6, and 8 Myr. 

\subsection{Observational completeness}

The last two columns in Table 3 show the NEOMOD2 prediction for the observational completeness of 
NEOs. To use the same magnitude system from which NEOMOD2 was derived, completeness is reported
relative to the MPC catalog from October 2022 (this defines the absolute magnitude system used
in this is work). The more recent MPC catalogs are (slightly) more complete as they include new NEO
discoveries since October 2022. Unfortunately, we cannot rigorously use these catalogs because the
absolute magnitudes reported for many NEOs and MBAs, for which we 
derived the CSS detection efficiencies, have changed: the $H$ magnitudes became on average  
fainter by a fraction of mag (Pravec et al. 2012). The shifting magnitudes mean that the actual 
number of NEOs brighter than $H$ (or larger than $D$) is lower ($N(H_2)$ values in Table 3
should be lower) than what we have inferred from the MPC catalog released in October
2022.\footnote{For example, Harris \& Chodas (2021) had 898 known NEOs with $H<17.75$ (MPC catalog
  from August 8, 2020), we have 854 known NEOs with $H<17.75$ (MPC catalog from October 19, 2022),
  and Harris \& Chodas (2023) have 805 known NEOs with $H<17.75$ (MPC catalog from March 13, 2023).
  As no individual asteroids were dropped from the MPC catalog, this means that nearly one hundred
  NEOs with estimated $H < 17.75$ in 2020 now have $H >17.75$. This is a dramatic shift.}
This issue, in itself, should not affect the completeness of the population reported in Table 3 (assuming
that NEOs and main belt asteroids are similarly affected). To rigorously test this, the magnitude 
system of NEOMOD would have to be updated to the new magnitudes, and this would essentially require 
to repeat many steps described in Sects. 2 and 3. We leave this for future updates. To approximately
align the estimated NEO population in this work with new MPC magnitudes, one can compare the number
of known NEOs in Table 3 ($N_{\rm MPC}(H_2)$) with the number of known NEOs with $H<H_2$ in any new
catalog, defining the ratio $f = N_{\rm MPC,new}(H_2)/N_{\rm MPC}(H_2)$, and apply it as a multiplication
factor to the NEOMOD estimate in Table 3 ($N(H_2)$), obtaining $N(H_2)_{\rm new}=f \times N(H_2)$. For example, if
$f=805/854=0.943$ for $H<17.75$ (Table 3 and MPC catalog from March 2023; Harris \& Chodas 2023),
then $N_{\rm new}(17.75) \simeq 0.943\times936=882$.

The results reported in Table 3 indicate that the population of small NEOs is largely incomplete
(Fig. \ref{compl}). For example, we find that the completeness for $H<22.75$ ($D>100$ m for $p_{\rm V}=0.14$) is 
$\simeq 26$\%. This compares reasonably well with HC21 who found a $\simeq 34$\% completeness
for $H<22.75$ (Harris \& Chodas 2023 estimated a $\simeq 40$\% completeness for $H<22.75$).
Our results start to diverge from HC21 and Harris \& Chodas (2023) for smaller NEOs. For the faintest 
magnitudes considered in our work, we find a $\simeq 0.3$\% completeness for $H<27.75$, whereas
HC21 and Harris \& Chodas (2023) reported only a $\simeq 0.09$\% completeness for $H<27.75$.
These differences are ultimately driven by the shallower absolute magnitude distribution that we
obtain here for $H>25$ (Fig. \ref{harris} and Sect. 4). 
 
Interestingly, the NEOMOD2 results suggest that even the population of bright NEOs could be significantly
incomplete. For example, the estimated completeness for $H<17.75$ is $91 \pm 4$\% (note that this 
is the completeness of the MPC catalog released in October 2022; the $H$ magnitudes updates in the new
catalogs complicate things). This would imply that 44--123 $H<17.75$ NEOs have yet to be 
discovered. The formal uncertainty of our estimate is relatively large. For comparison, HC21
and Harris \& Chodas (2023) used the redetection method to estimate the completeness of $H<17.75$ NEOs
at $\simeq 96$\%, which is just about one sigma above our estimate. The redetection method may provide
a more accurate completeness estimate for these bright NEOs for which the redetection statistics
is good. Related to that, it would be worthwhile to quantify various uncertainties of the redetection
method and their impact on the completeness estimate. Seven $H<17.75$ NEOs were discovered in the
past two years (2022-2023): 2022 KL8, 2023 HQ2, 2022 AP7, 2023 PS2, 2022 QK204, 2023 GZ1, and 2022
RX3. Many of these have large inclinations (five have $i>30^\circ$) and/or large semimajor axes
(five have $a>2.8$ au). At this rate, it could take over a decade to find 99\% of all $H<17.75$
NEOs ($a<4.2$ au).

To understand where the bright NEOs may be hiding, we generated a large sample of bright NEOs from the NEOMOD 
Simulator and ran them through the CSS detection pipeline. We used the G96 observations from 
2005-2022 and 703 observations from 2005-2012. This cumulatively corresponds to 28 years of NEO
observations from the northern hemisphere. We found that $\simeq 4$\% of bright NEOs do not appear 
in any frame taken by CSS and would thus avoid detection. Most of these objects have $a \lesssim 
1.2$ au and avoid detection due to the synodic effect, or have the argument of perihelion 
$\omega \sim 90^\circ$ -- and therefore appear in the southern hemisphere near opposition. Two 
ATLAS NEO-survey telescopes with large FoVs just started operations from the southern hemisphere 
(Chile and Southern Africa). By simulating their detection capabilities (Deienno et al. 2023),
we find that both of these telescopes should be very effective in detecting bright NEOs that 
escape detections from the northern hemisphere. Indeed, W68 (Chile) has recently a discovery of 
2022 RX3, a potentially hazardous NEO with $H=17.7$.  

\subsection{Collisional evolution of small NEOs}

We used the \"Opik formalism (Bottke et al. 1994) to estimate the collisional probabilities and 
velocities among NEOs, and between NEOs and main belt asteroids (MBAs). The intrinsic probability for 
collisions among NEOs is relatively high, $P_{\rm i} \simeq 6.5 \times 10^{-18}$ km$^{-2}$ yr$^{-1}$,
but the population of NEOs is much smaller than MBAs; collisions among NEOs can therefore be neglected. 
The probabilities and velocities for collisions between NEOs and
MBAs are $P_{\rm i} \simeq 2.6 \times 10^{-18}$ km$^{-2}$ yr$^{-1}$ and $V_{\rm i} \simeq 11.6$ km/s.
The impact speeds are therefore $\simeq 2$ times higher than in the case of collisions among MBAs.
Taking this into account we estimate that the collisional lifetime of NEOs should be $\sim 3$ times
shorter, on average, than the collisional lifetime of MBAs. For MBAs, Bottke et al. (2005) reported that
the average collisional lifetime is $\sim 30$ Myr for $D \simeq 10$ m. This allows us to estimate
that the collisional lifetime of $D \simeq 10$ m NEOs is $\sim 10$ Myr, only slightly longer than
the dynamical lifetime of NEOs produced from the $\nu_6$ resonance (Table 5 in Paper I; other 
resonances give much shorter dynamical lifetimes). This means that it may be justified, but 
barely so, to neglect the collisional evolution of NEOs for $D > 10$ m. Conversely, for $D < 10$ m,
the collisional lifetime of NEOs would have to be taken into account. For reference, here we 
estimate that the average collisional lifetime of $D \simeq 1$ m NEO is $\sim 5$ Myr.   

\subsection{Distribution of NEO obliquities}

La Spina et al. (2004) reported a $\sim 2$:1 preference for retrograde rotation among large NEOs
($D \gtrsim 1$ km). They interpreted this result in the context of the NEO model from Bottke et al. (2002).
In the Bottke et al. model, the $\nu_6$ resonance contributes to $\simeq 37$\% of NEOs. To reach $\nu_6$,
a main belt asteroid must have a retrograde rotation and drift inward; this implies that the 
$\nu_6$ resonance should produce predominantly retrograde NEOs. Other important sources of NEOs, 
including the 3:1 resonance and weak resonances in the inner belt, can be reached from both sides, and this 
implies that they should be producing and roughly equal share of prograde and retrograde NEOs.
La Spina et al. (2004) therefore found from this argument that the ratio of retrograde to prograde
NEOs should be $(37+63/2)$:$63/2$ or $\sim 2$:1, in good agreement with observations.      

NEOMOD2 indicates a much smaller contribution of $\nu_6$ resonance to large NEOs: $\alpha_{\nu 6}=0.06 
\pm 0.03$ for $H\simeq15$. If this is correct, the contribution of the $\nu_6$ resonance to large
retrograde NEOs would be minimal. We therefore suggest that the preference for retrograde rotation
of large NEOs is probably related to something else. There are at least two possibilities: 

(1) We find that the number of $H<18$ MBAs on the sunward side of the 3:1 resonance is significantly 
lower (by $\sim 50$\%) than on the opposite side. This asymmetry, which favors generation of retrograde 
NEOs from 3:1, is contributed by asteroid families (Nesvorn\'y et al. 2015). 

(2) \v{D}urech \& Hanu\v{s} (2023) obtained the distribution of $D>1$ km MBAs from a Gaia-DR3 data analysis 
(Gaia Collaboration et al. 2023). They showed that retrograde MBAs often have the obliquity $\theta 
\simeq 180^\circ$, most likely because they reached the terminal state of the 
YORP evolution (Vokrouhlick\'y et al. 2015). The prograde MBAs, however, show a broader distribution 
of obliquities (roughly $0 < \theta \lesssim 60^\circ$). This presumably happens because prograde 
rotators can be captured in spin-orbit resonances that can prevent them from reaching $\theta \simeq 0$
(Vokrouhlick\'y et al. 2003). All this means that the retrograde MBAs should have, on average, faster 
Yarkovsky drift rates (the Yarkovsky drift rate scales with $\cos \theta$; Vokrouhlick\'y et al. 2015) 
than prograde MBAs; they more likely reach resonances and evolve onto NEO orbits.\footnote{It needs 
to be demonstrated whether the asymmetric feeding of resonances by faster drifting 
retrograde MBAs can remain in a steady state. Without a distant source, the retrograde MBAs on 
the outer side of the 3:1 resonance ($a>2.5$ au) would end up evolving into the resonance. Their 
number density in a narrow strip near the 3:1 resonance would decrease, and this would affect the 
feeding rate. In reality, however, the number density of MBAs on the outer side of the 3:1 resonance 
is larger than on the sunward side (see item (1) above).}
 
Asymmetric feeding of the 3:1 and other strong resonances could provide a possible explanation for the 
preference for retrograde rotation among large NEOs
(La Spina et al. 2004). Farnocchia et al. (2013) analyzed obliquities of small, sub-km NEOs and found
that $81 \pm 8$\% have retrograde rotation (i.e., roughly a 4:1 preference for retrograde rotation).
The increasing share of retrograde rotators among smaller NEO is likely related the fact that 
the $\nu_6$ resonance contribution to the NEO population increases for small bodies. For example, 
for $D \simeq 0.1$ km ($H \simeq 22.75$ for $p_{\rm v}=0.14$), the $\nu_6$ contribution is $\simeq 
40$\%, which is already similar to the $\nu_6$ contribution adopted in La Spina et al. (2004). 
This presumably could, when combined with obliquity distribution differences discussed above, explain 
the 4:1 preference for retrograde rotation among small NEOs (Farnocchia et al. 2013).

\section{Summary}

The main results of this work are summarized as follows.

\begin{description}
\item (1) We updated the previous NEO model. NEOMOD v2.0 is based on numerical integrations 
  of bodies from 12 sources (11 main-belt sources and comets). A flexible method to accurately calculate
  biases of NEO surveys was applied to the Catalina Sky Survey (CSS) observations from 2013 to 2022, when 
  CSS detected $\simeq 14,$000 unique NEOs (this can be compared to only $\simeq 4,$500 unique NEOs detected 
  by CSS from 2005 to 2012 (Paper I, Granvik et al. 2018). The \texttt{MultiNest} code (Feroz \& Hobson 
  2008, Feroz et al. 2009) was used to optimize the model fit to CSS detections. 
\item (2) The best-fit orbital and absolute magnitude NEO model is available via the NEOMOD 
  Simulator,\footnote{\url{https://www.boulder.swri.edu/\~{}davidn/NEOMOD\_Simulator} and GitHub.} a code 
  that can be used to generate user-defined NEO samples from the model. Researchers interested in the probability
  that a specific NEO evolved from a particular source can obtain this information from the ASCII table
  that is available along with the Simulator.
\item (3) We confirm that the sampling of main-belt sources by NEOs is {\it size-dependent} with the 
  $\nu_6$ and 3:1 resonances contributing $\sim 30$\% of NEOs with $H \sim 15$, and $\sim 90$\% of 
  NEOs with $H \sim 28$. This trend most likely arises from how the small and large main-belt asteroids 
  reach the source regions. We confirm the size-dependent disruption of NEOs reported in Granvik et al. 
  (2016) and Paper I. As a consequence of the size-dependent sampling and disruption, small and large NEOs 
  have different orbital distributions.
\item (4) We found a shallower absolute magnitude distribution for $25<H<28$ and smaller number of NEOs 
  with $H<28$ than Harris \& Chodas (2021). This may point to some problem with the detection efficiency
  of CSS. Alternatively, some of the assumptions in Harris \& Chodas (2021) may not be quite right. 
  When tidal disruption is ignored, the average time between terrestrial impacts of $D > 10$ m bolides 
  is found here to be 29 yr -- $\simeq 1.5$ and $\simeq 3$ times longer than the nominal estimates from 
  Harris \& Chodas (2021) and Brown et al. (2002, 2013). See item (6) below for the results with tidal 
  disruption. 
\item (5) We estimate $936 \pm 29$ NEOs with $H<17.75$ ($D>1$ km for $p_{\rm V}=0.14$) and $a<4.2$ au. 
  With 854 known $H<17.75$ NEOs (as of October 2022), the NEO population with $H<17.75$ is 87--95\% complete 
 (1$\sigma$ interval). Many of the yet-to-be-detected bright NEOs should have large orbital inclinations and/or 
  large semimajor axes. The hemispheric bias will be reduced as two ATLAS telescopes continue to operate 
  from the south hemisphere. The known NEO population with $H<22$ ($D>140$~m for $p_{\rm V}=0.14$) is only 47-49\% 
  complete. 
\item (6) The excess of CSS NEO detections for $1<a<1.6$ au, $q \simeq 1$ au, $i \lesssim 10^\circ$ 
  and $25<H<28$ (Figs. \ref{faint} and \ref{excess}) is attributed to tidal disruption of larger NEOs during close 
  encounters with the Earth. The orbital fit significantly improves in a model where tidal disruption is
  (approximately) accounted for. With tidal disruption, the average time between terrestrial impacts of 
  $D > 10$ m bolides is found to be $\simeq 17$~yr. Tidal disruption could also help to explain the PM excess 
  of meteorite falls and differences in lunar and Martian crater size distributions.  
\item (7) For $H \sim 28$, the Mars-to-Moon impact ratio, normalized to the unit surface area, is found to be 
  $R_{\rm b}=0.8$ in the model without tidal disruption and $R_{\rm b}=0.5$ in the model with tidal disruption.
  Previous works used $R_{\rm b}=2.6$ for all asteroid impactor sizes (Hartmann 2005, Marchi 2021) to apply 
  the lunar crater chronology to Mars; this may be incorrect, especially for small impactors. The trend of 
  decreasing $R_{\rm b}$ for smaller impactors is consistent with the results reported in Paper I, where we 
  found $R_{\rm b}=1.2$ for $H \sim 25$.
\item (8) We suggest that the distribution of obliquities of large NEOs, which shows a $\sim 2$:1 preference 
  for retrograde rotation (La Spina et al. 2004), may be related to (on average) faster Yarkovsky drift rates 
  of retrograde main belt asteroids (given that their obliquities are more tightly clumped near 180$^\circ$,
  \v{D}urech \& Hanu\v{s} (2023); Sect. 7.7), and/or to the asymmetric distribution of main belt asteroids 
  around source resonances (e.g., 3:1; Sect 7.7).  
  The larger share of retrograde rotators among smaller NEOs 
  (Farnocchia et al. 2013) is likely related the fact that the $\nu_6$ resonance contribution increases 
  for small NEOs (the $\nu_6$ resonance can only be reached by sunward-drifting bodies with $\theta>90^\circ$).    
\item (9) The impact probability of a $H<22$ ($D>140$ m for $p_{\rm V}=0.14$) object on the Earth in this 
  millennium is estimated to be $\simeq 4.5$\%. 
\end{description}

\acknowledgements
The simulations were performed on the NASA Pleiades Supercomputer. We thank the NASA NAS computing division 
for continued support. The work of DN, RD, WFB and DF was supported by the NASA Planetary Defense Coordination 
Office project ``Constructing a New Model of the Near-Earth Object Population''. The work of SN, SRC, PWC and DF 
was conducted at the Jet Propulsion Laboratory, California Institute of Technology, under a contract with 
the National Aeronautics and Space Administration. DV acknowledges support from the grant 21-11058S of 
the Czech Science Foundation. We thank Alan Harris and the anonymous reviewer for helpful comments. Alan Harris 
graciously provided data from Harris \& Chodas (2023) for Fig. 16.

\clearpage

\begin{table}
\centering
{
\begin{tabular}{lrr}
\hline \hline 
                  & CSS1  & CSS2 \\  
\hline
$\epsilon_0$      & 0.983 & 0.952     \\
$V'_0$             &  -6.0 & -2.65    \\ 
$q_{\rm V}$        &  23.87 &  6.64     \\
$V'_{\rm lim}$       &  0.475 &  0.170     \\
$V_{\rm wid}$       &  0.180 &  0.175     \\
$\alpha$          &  1.140 &  1.151      \\
\hline \hline
\end{tabular}
}
\caption{Global photometric parameters of CSS1 and CSS2. See Sect. 2.1 and Eq. (\ref{photo})
  for the definition of these parameters: $\epsilon_0$ defines the detection efficiency
  for bright apparent magnitudes, $V'_0$
and $q_V$ are parameters of the quadratic term that improve the behavior of the
analytic fit for $V'<V'_{\rm lim}$, $V'_{\rm lim}$ is the apparent magnitude where
the detection efficiency drops, $V_{\rm wid}$ defines how fast it drops, and $\alpha$
improves the behavior of the analytic fit for $V'>V'_{\rm lim}$.
CSS2 has a lower value of global $V'_{\rm lim}$ than CSS1. 
This means that, for CSS2, $V_{\rm 3rd}$ is a better proxy for where the photometric detection 
efficiency drops.  The large value
of CSS1's $q_V$ reduces the importance of the quadratic term; this term is more
important for CSS2. The values reported here were computed for the apparent motion $0.12<w<1$
deg/day.  }
\end{table}

\begin{table}
\centering
{ \small
\begin{tabular}{lrrrrr}
\hline \hline
label & parameter        & median & $-\sigma$ & $+\sigma$ & limit \\                      
\hline
\multicolumn{6}{c}{$\alpha$'s for $H=15$}                    \\
(1) & $\nu_6$          & 0.060   & 0.003    & 0.003 & --     \\ 
(2) & 3:1              & 0.277   & 0.028    & 0.028 & --     \\
(3) & 5:2              & 0.073   & 0.018    & 0.019 & --     \\ 
(4) & 7:3              & 0.007   & 0.005    & 0.007 & 0.010  \\ 
(5) & 8:3              & 0.103   & 0.013    & 0.013 & --     \\ 
(6) & 9:4              & 0.008   & 0.005    & 0.011 & 0.012  \\                 
(7) & 11:5             & 0.076   & 0.014    & 0.015 & --   \\   
(8) & 2:1              & 0.039   & 0.005    & 0.006 & --   \\   
(9) & inner weak       & 0.183   & 0.026    & 0.025 & --   \\  
(10) & Hungarias        & 0.063   & 0.013    & 0.012 & --   \\ 
(11) & Phocaeas         & 0.094   & 0.010    & 0.010 & --   \\ 
 -- & JFCs             & 0.012   & 0.006    & 0.007  & 0.016  \\ 
\hline
\multicolumn{6}{c}{$\alpha$'s for $H=28$}                \\
(12) & $\nu_6$          & 0.595   & 0.024    & 0.022 & --    \\ 
(13) & 3:1              & 0.313   & 0.020    & 0.020 & --    \\
(14) & 5:2              & 0.019   & 0.009    & 0.010 & --    \\ 
(15) & 7:3              & 0.003   & 0.002    & 0.004 & 0.005 \\ 
(16) & 8:3              & 0.004   & 0.003    & 0.006 & 0.006   \\ 
(17) & 9:4              & 0.003   & 0.002    & 0.004 & 0.005    \\                 
(18) & 11:5             & 0.004   & 0.003    & 0.006 & 0.006    \\   
(19) & 2:1              & 0.001   & 0.001    & 0.002 & 0.002    \\   
(20) & inner weak       & 0.008   & 0.006    & 0.013 & 0.014   \\  
(21) & Hungarias        & 0.020   & 0.014    & 0.020 & 0.029     \\ 
(22) & Phocaeas         & 0.003   & 0.002    & 0.004 & 0.004  \\ 
-- & JFCs               & 0.014   & 0.007    & 0.008 & 0.018  \\ 
\hline
\multicolumn{6}{c}{$H$ {\it distribution}} \\
(23) & $N_{\rm ref}$      & 926     & 29       & 29    & --     \\  
(24) & $\gamma_2$        & 0.393   & 0.013    & 0.014 & --     \\
(25) & $\gamma_3$        & 0.363   & 0.006    & 0.006 & --      \\
(26) & $\gamma_4$        & 0.313   & 0.003    & 0.003 & --      \\
(27) & $\gamma_5$        & 0.522   & 0.006    & 0.006 & --       \\
(28) & $\gamma_6$        & 0.506   & 0.005    & 0.005 & --       \\
\hline
\multicolumn{6}{c}{{\it Disruption parameters}}      \\
(29) & $q^*_0$          &  0.132   & 0.003    & 0.002  & --     \\ 
(30) & $\delta q^*$     &  0.031  & 0.001    & 0.001   & --    \\  
\hline \hline
\end{tabular}
}
\caption{The median and uncertainities of our base model parameters.
The uncertainties reported here were obtained from the posterior distribution 
produced by \texttt{MultiNest}. They do not account for uncertainties of the CSS detection 
efficiency. For parameters, for which the posterior distribution 
peaks near zero, the last column reports the upper limit (68.3\% of posteriors fall between
zero and that limit).}
\end{table}

\begin{table}
\centering
{ \footnotesize
\begin{tabular}{rrrrrrrrrr}
\hline \hline                     
  $H_1$  &  $H_2$ &   ${\rm d}N$  &  $N(H_2)$  &  $N_{\rm HC}(H_2)$ &  $N_{\rm min}(H_2)$ & $N_{\rm max}(H_2)$ &  $N_{\rm MPC}(H_2)$ 
& Compl. & Range \\
\hline

 15.25 & 15.75  & 61.2      & 130.      & 136.      & 124.      & 137.       & 123.  & 95\% & (90-99) \\
 15.75 & 16.25  & 104.      & 234.      & 235.      & 219.      & 250.       & 210.  & 90\% & (84-96) \\
 16.25 & 16.75  & 156.      & 390.      & 398.      & 365.      & 416.       & 361.  & 93\% & (87-99) \\
 16.75 & 17.25  & 218.      & 608.      & 621.      & 579.      & 639.       & 562.  & 92\% & (88-97) \\
 17.25 & 17.75  & 328.      & 936.      & 940.      & 898.      & 977.       & 854.  & 91\% & (87-95) \\
 17.75 & 18.25  & 513.     & 0.145E4 & 0.147E4 & 0.140E4 & 0.151E4  & 1325.  & 91\% & (88-95) \\
 18.25 & 18.75  & 790.     & 0.224E4 & 0.221E4 & 0.217E4 & 0.232E4  & 2022.  & 90\% & (87-93) \\
 18.75 & 19.25 & 0.117E4 & 0.341E4 & 0.323E4 & 0.331E4 & 0.350E4  & 2897.  & 85\% & (83-88) \\
 19.25 & 19.75 & 0.164E4 & 0.505E4 & 0.463E4 & 0.492E4 & 0.517E4  & 4021.  & 80\% & (78-82) \\
 19.75 & 20.25 & 0.216E4 & 0.721E4 & 0.642E4 & 0.703E4 & 0.737E4  & 5281.  & 73\% & (72-75) \\
 20.25 & 20.75 & 0.272E4 & 0.992E4 & 0.873E4 & 0.970E4 & 0.101E5  & 6636.  & 67\% & (66-68) \\
 20.75 & 21.25 & 0.350E4 & 0.134E5 & 0.118E5 & 0.131E5 & 0.137E5  & 8076.  & 60\% & (59-60) \\
 21.25 & 21.75 & 0.471E4 & 0.181E5 & 0.159E5 & 0.178E5 & 0.185E5  & 9480.  & 52\% & (51-53) \\
 21.75 & 22.25 & 0.673E4 & 0.249E5 & 0.217E5 & 0.244E5 & 0.254E5 & 10865.  & 44\% & (43-45) \\
 22.25 & 22.75 & 0.104E5 & 0.353E5 & 0.314E5 & 0.345E5 & 0.360E5 & 12309.  & 35\% & (34-36) \\
 22.75 & 23.25 & 0.173E5 & 0.525E5 & 0.476E5 & 0.514E5 & 0.536E5 & 13862.  & 26\% &   -- \\
 23.25 & 23.75 & 0.311E5 & 0.836E5 & 0.826E5 & 0.818E5 & 0.853E5 & 15673.  & 19\% &   -- \\
 23.75 & 24.25 & 0.608E5 & 0.144E6 & 0.153E6 & 0.142E6 & 0.147E6 & 17622.  & 12\% &   -- \\
 24.25 & 24.75 & 0.121E6 & 0.266E6 & 0.313E6 & 0.260E6 & 0.272E6 & 19709. & 7.4\% &   -- \\
 24.75 & 25.25 & 0.229E6 & 0.494E6 & 0.641E6 & 0.482E6 & 0.506E6 & 21724. & 4.4\% &   -- \\
 25.25 & 25.75 & 0.411E6 & 0.905E6 & 0.130E7 & 0.882E6 & 0.928E6 & 23636. & 2.6\% &   -- \\
 25.75 & 26.25 & 0.728E6 & 0.163E7 & 0.241E7 & 0.159E7 & 0.168E7 & 25337. & 1.6\% &   -- \\
 26.25 & 26.75 & 0.129E7 & 0.292E7 & 0.481E7 & 0.284E7 & 0.300E7 & 26728. & 0.9\% &   -- \\
 26.75 & 27.25 & 0.225E7 & 0.517E7 & 0.108E8 & 0.500E7 & 0.534E7 & 27849. & 0.5\% &   -- \\
 27.25 & 27.75 & 0.395E7 & 0.912E7 & 0.244E8 & 0.875E7 & 0.949E7 & 28653. & 0.3\% &   -- \\
\hline \hline
\end{tabular}
}
\caption{The absolute magnitude distribution and completeness of the NEO population. The columns are: 
the lower limit of a magnitude bin ($H_1$), upper limit of a magnitude bin ($H_2$), NEOMOD estimate 
of the number of NEOs between $H_1$ and $H_2$ (${\rm d}N$), NEOMOD estimate of the number of NEOs 
with $H<H_2$ ($N(H_2)$), Harris \& Chodas (2021) estimate of $N(H_2)$ ($N_{\rm HC}(H_2)$), NEOMOD estimate 
of $N(H_2)$ minus $1\sigma$ ($N_{\rm min}(H_2)$), NEOMOD estimate of $N(H_2)$ plus $1\sigma$ ($N_{\rm max}(H_2)$),
number of NEOs with $H<H_2$ in the MPC catalog from October 2022 ($N_{\rm MPC}(H_2)$), completeness 
defined as $N_{\rm MPC}(H_2)/N(H_2)$, and 1$\sigma$ completeness range ($<1$\% uncertainties not listed).}
\end{table}

\clearpage
\begin{figure}
\epsscale{1.0}
\plotone{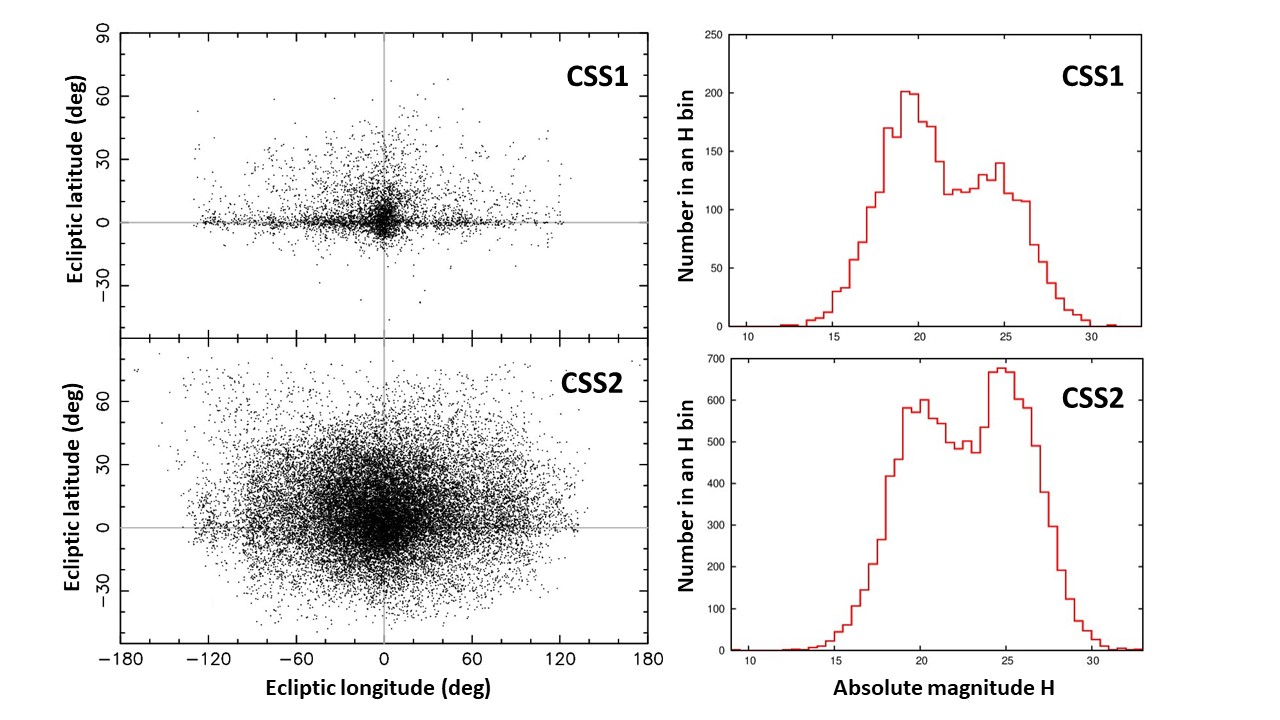}
\caption{NEOs detected by CSS1 (2013-2016; upper panels) and CSS2 (2016-2022; lower panels).
The plots on the left show the ecliptic coordinates of detected objects. The plots on the right
show their absolute magnitude distributions.}
\label{detections}
\end{figure}


\clearpage
\begin{figure}
\epsscale{1.5}
\hspace*{-4cm}\plotone{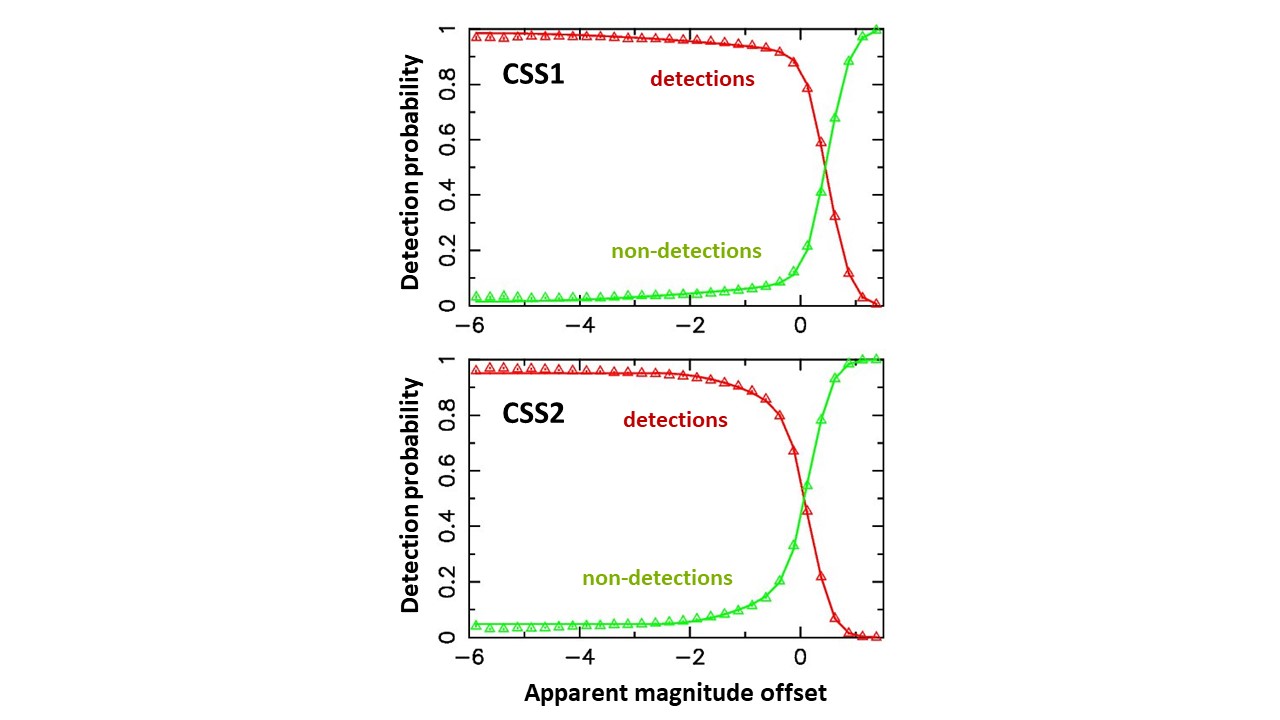}
\caption{Global photometric sensitivities of CSS1 (top panel) and CSS2 (bottom panel). The red 
triangles show the binned detection probability, $\epsilon(V')=N_{\rm det}(V')/N_{\rm all}(V')$
(Eq. \ref{photo}), as a function of the apparent visual magnitude offset $V'$ (Sect. 2.1). The green triangles show the 
probability of non-detection, $1-\epsilon(V')$. The red and green lines show the best fits 
to the binned data using the functional dependence given in Eq.~(\ref{photo}).}
\label{css}
\end{figure}

\clearpage
\begin{figure}
\epsscale{1.5}
\hspace*{-4cm}\plotone{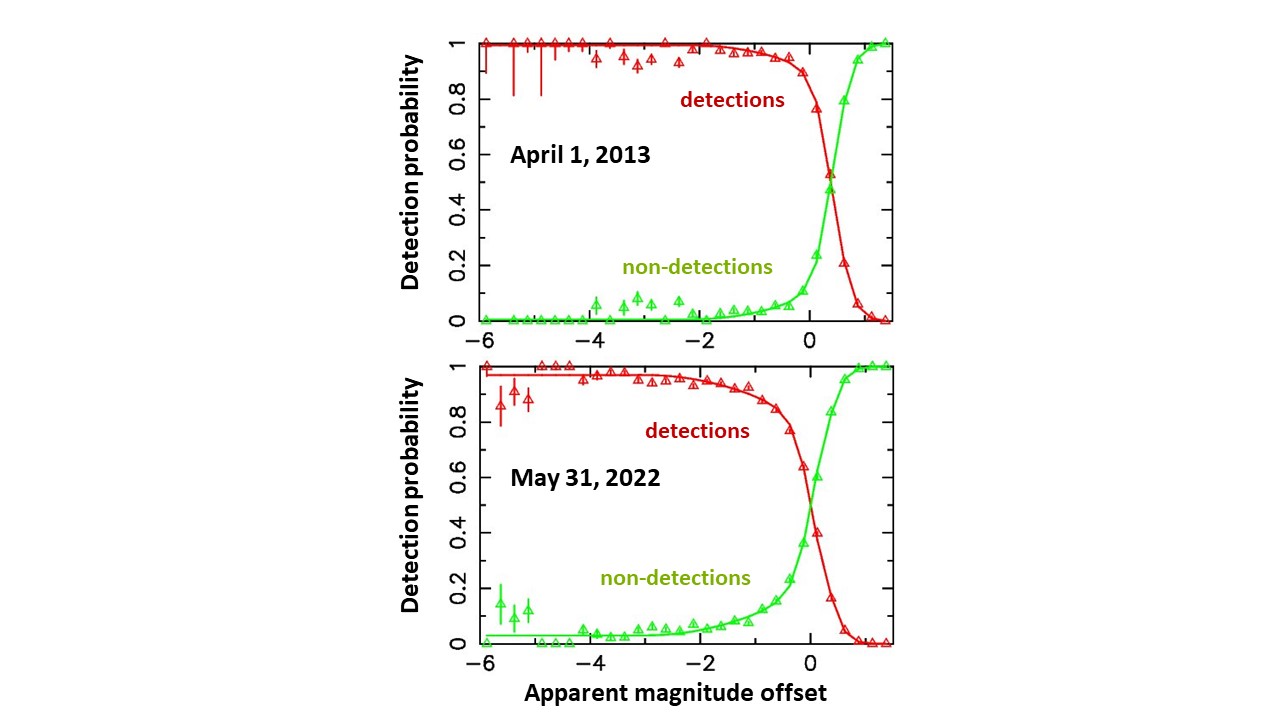}
\caption{Examples of nightly photometric detection probabilities for CSS1 (April 1, 2013; top
panel) and CSS2 (May 31, 2022; bottom panel). See the caption of Fig. \ref{css} for the description
of symbols and lines. The error bars were estimated adopting the Poisson statistics.}
\label{night}
\end{figure}

\clearpage
\begin{figure}
\epsscale{1.0}
\plotone{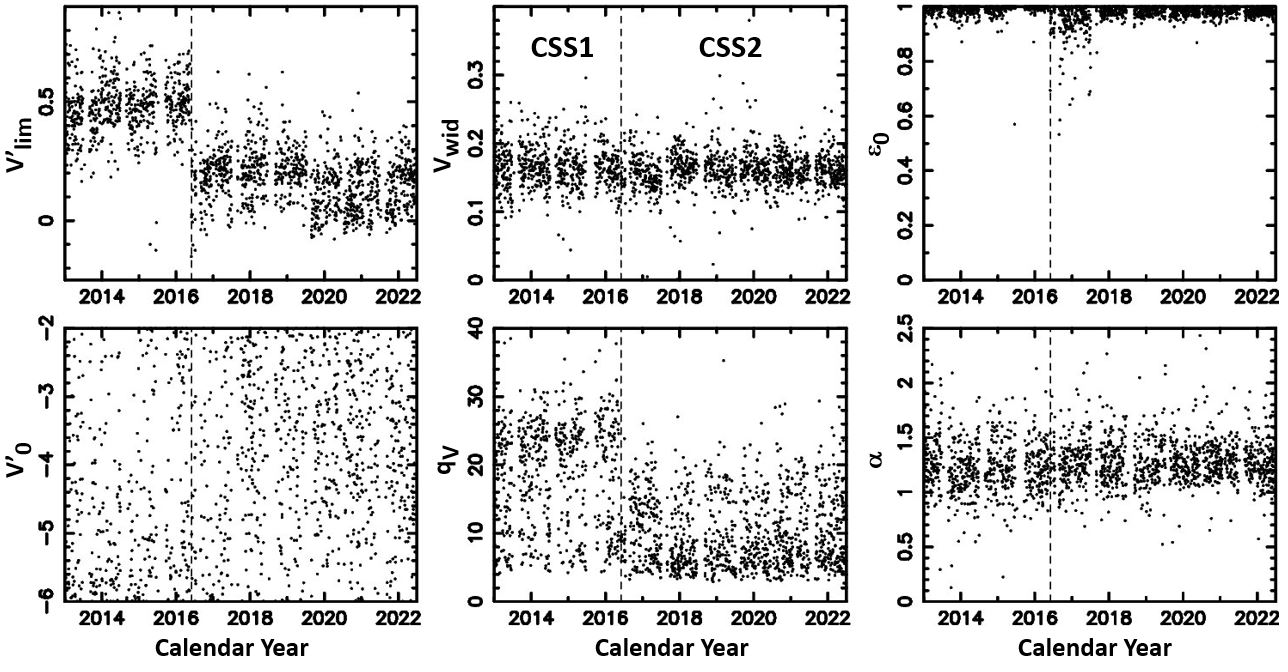}
\caption{Variation of six photometric parameters (Eq. \ref{photo}) derived on a nightly basis,
for the whole duration of new CSS (January 2013 to June 2022). The G96 telescope was upgraded 
in May 2016 (vertical dashed lines).}
\label{photop}
\end{figure}

\clearpage
\begin{figure}
\epsscale{1.5}
\hspace*{-4cm}\plotone{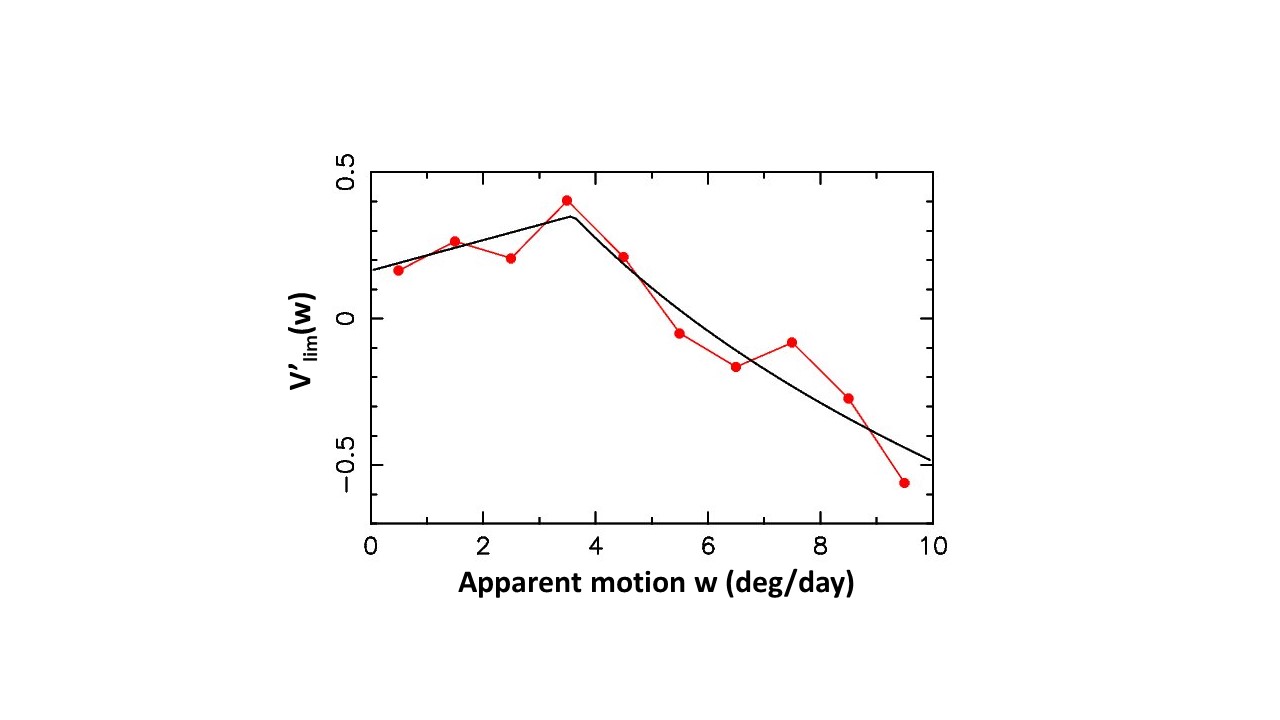}
\caption{The dependence of the transition magnitude $V'_{\rm lim}$ on the asteroid's apparent motion 
$w$. The red line and dots show $V'_{\rm lim}(w)$ obtained from the Simplex fit to all new CSS 
observations. The black line is the analytic fit with the functional form described in the 
main text (Eqs. (3) and (4)).}  
\label{trail}
\end{figure}


\clearpage
\begin{figure}
\epsscale{1.0}
\plotone{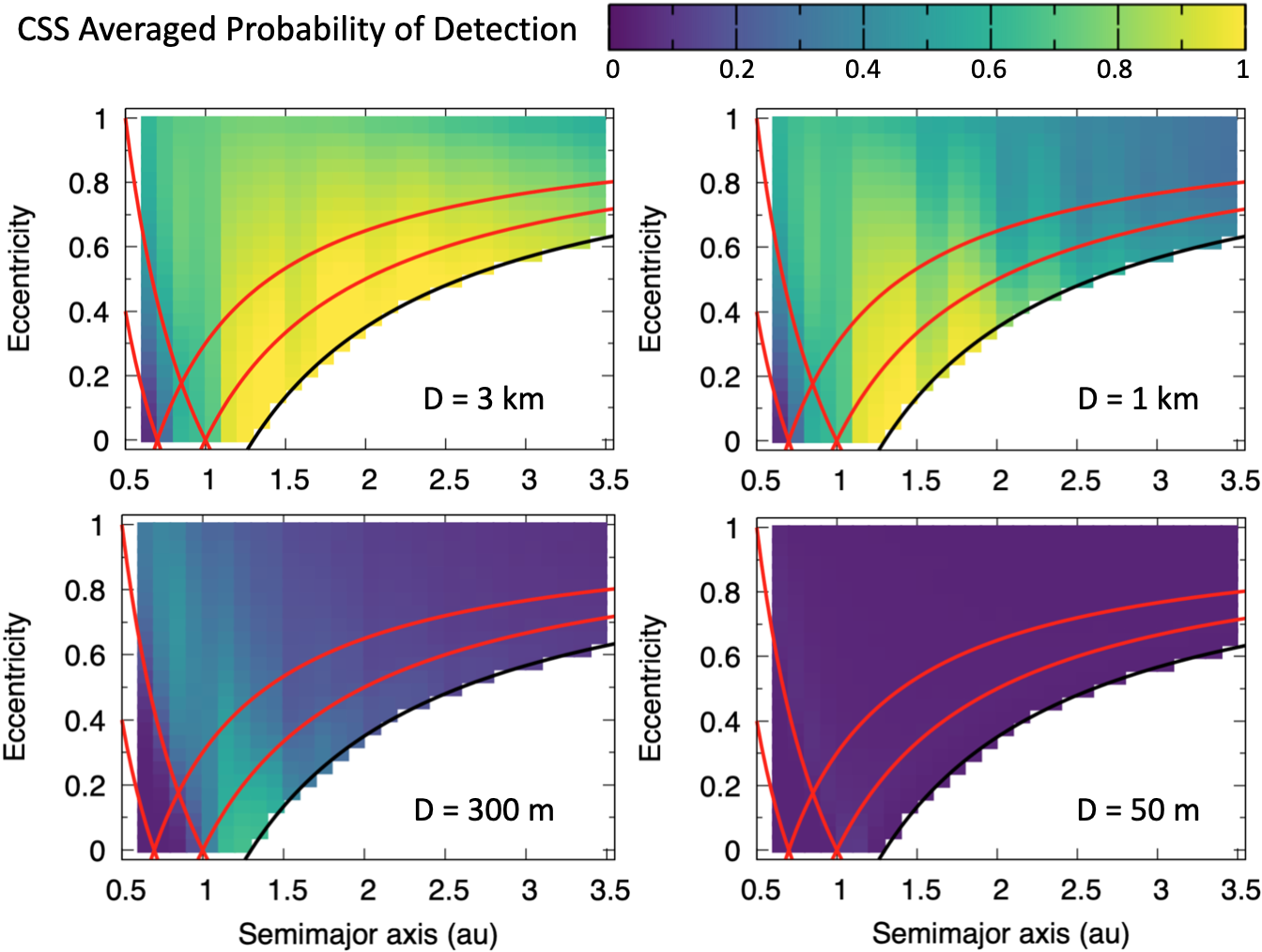}
\caption{The CSS2 detection probability (Eq. \ref{calp}) as a function of orbital elements for four different 
  absolute magnitude values.  From top-left to bottom-right, we plot ${\cal P}(a,e,i,H)$ for $H$ corresponding
  to objects with $D=3$ km, 1 km, 300 m and 50 m (for the reference albedo $p_{\rm V}=0.14$). The detection
  probability was averaged over all inclinations bins. The vertical strips, with ${\cal P}$ going up and down
  as a function of NEO's semimajor axis, are a consequence of the synodic effect (see discussion in Paper I).
  The red lines show borders of the orbital domain where orbits can have close encounters with Earth and
  Venus.}
\label{bias2}
\end{figure}

\clearpage
\begin{figure}
\epsscale{0.8}
\plotone{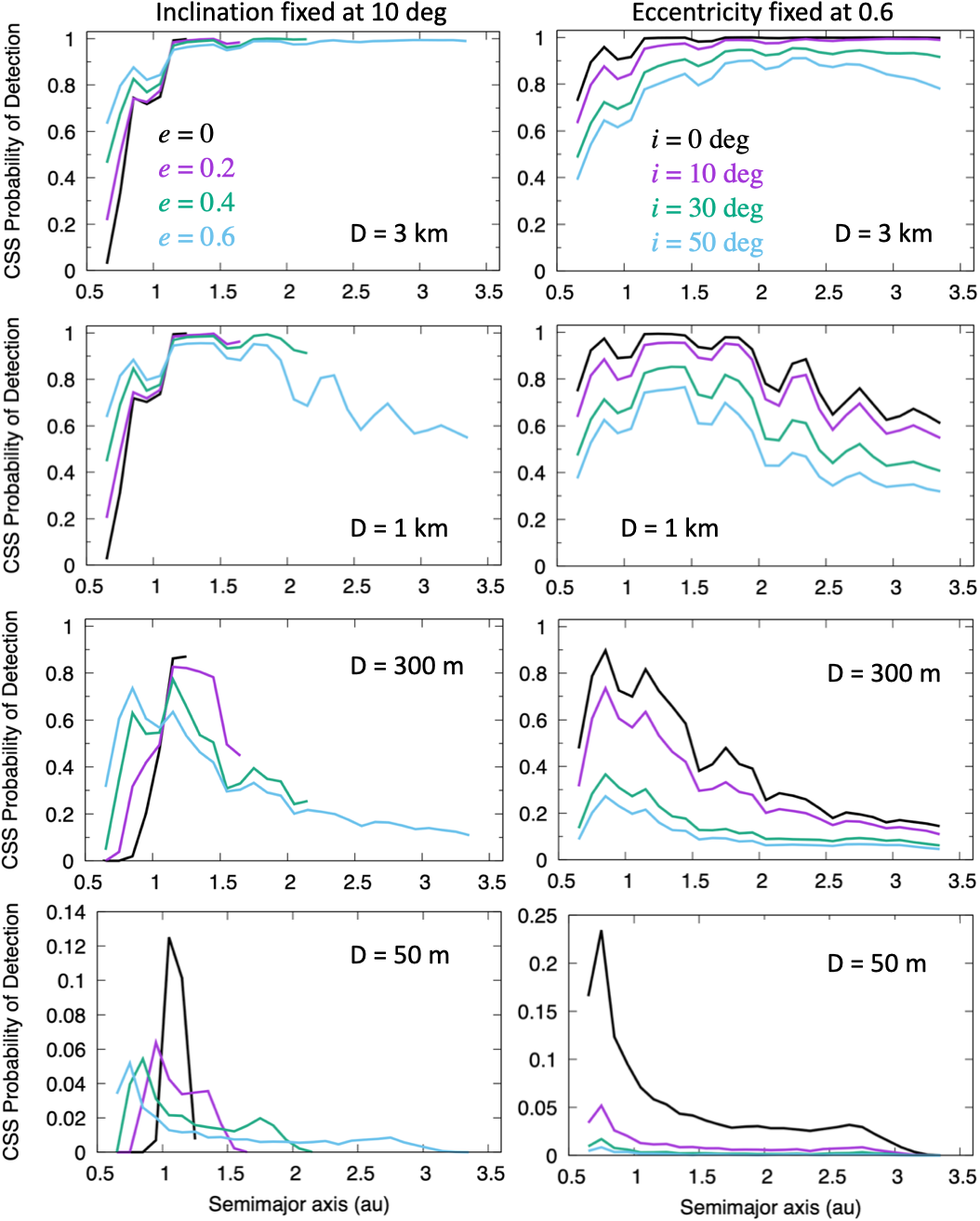}
\caption{The CSS2 detection probability (Eq. \ref{calp}) as a function of orbital elements for four 
different absolute magnitude values.  From top to bottom, we plot ${\cal P}(a,e,i,H)$ for 
$H$ corresponding to objects with $D=3$ km, 1 km, 300 m and 50 m (for the reference albedo $p_{\rm V}=0.14$).
The plots in the left column show ${\cal P}$ for the fixed orbital inclination ($i=10^\circ$) and several
eccentricity values. The plots on the right show ${\cal P}$ for $e=0.6$ and several inclination values.
The detection probability was computed for orbits with $q<1.3$ au.}
\label{bias3}
\end{figure}


\clearpage
\begin{figure}
\epsscale{0.6}
\plotone{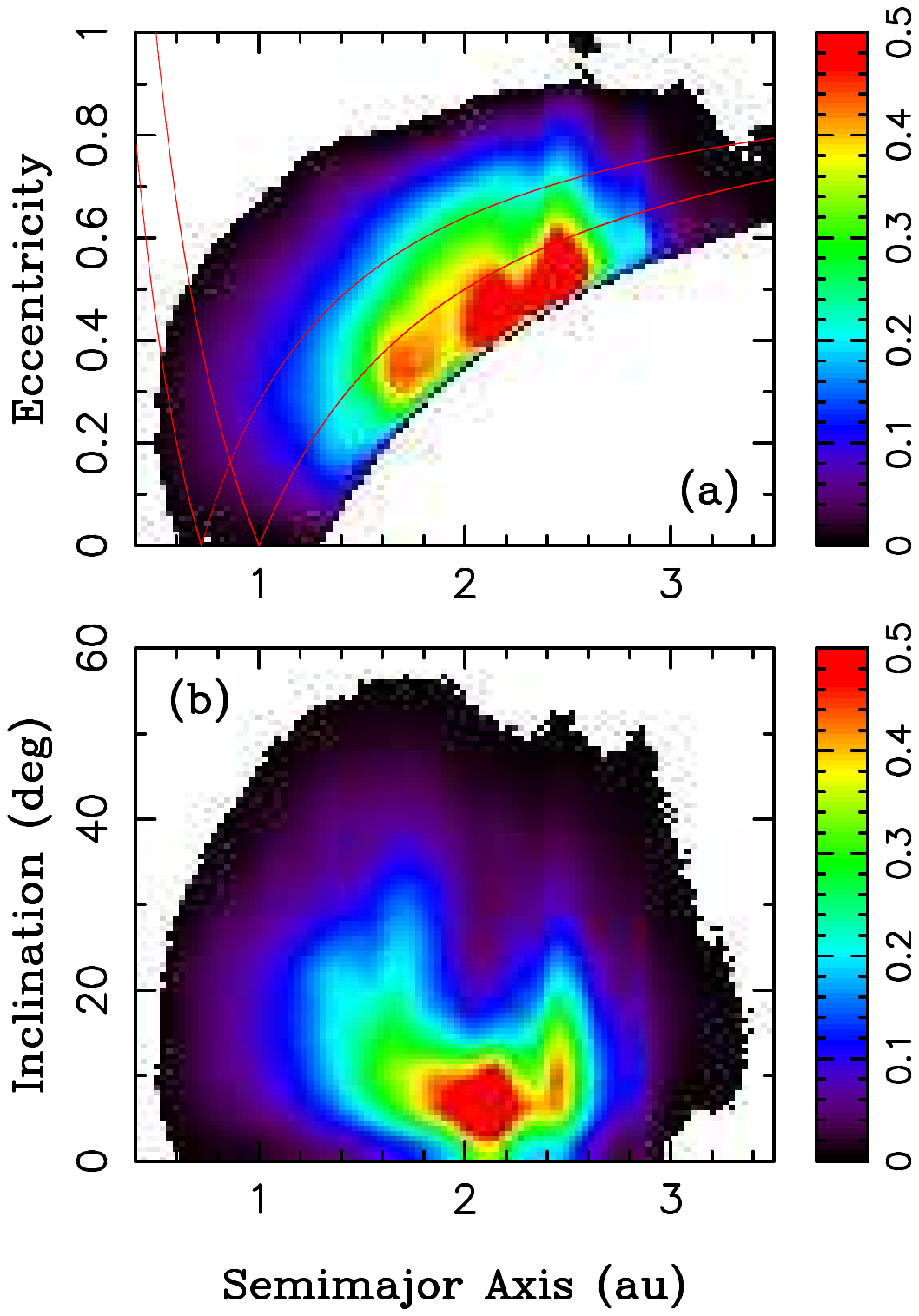}
\caption{The orbital distribution of NEOs from our {\it intrinsic} (debiased) best-fit model ${\cal M}$. 
We used the NEOMOD Simulator (Sect. 4) and generated $1.1\times10^6$ NEOs with $15<H<28$. The distribution 
was marginalized over absolute magnitude and binned using 100 bins in each orbital element 
($0.4<a<3.5$ au, $e<1$ and $i<60^\circ$). Warmer colors correspond to orbits where NEOs are more likely
to spend time. The red lines show borders of the orbital domain where orbits can have close encounters
with Earth and Venus.}
\label{unb}
\end{figure}

\clearpage
\begin{figure}
\epsscale{1.35}
\hspace*{-3.cm}\plotone{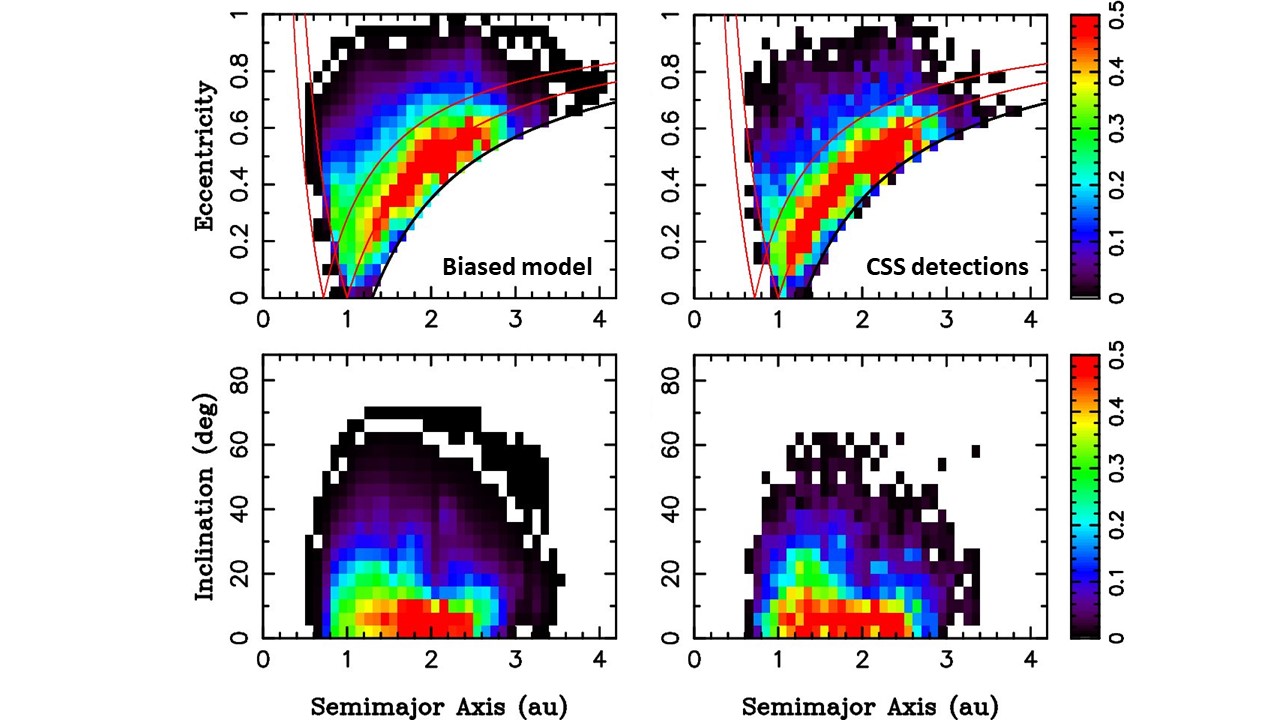}
\caption{The orbital distribution of NEOs from our {\it biased} best-fit model ${\cal M}_{\rm b}$ (left 
panels) and the CSS NEO detections (right panels). The model distribution for $15<H<28$ was marginalized 
over absolute magnitude and binned with the standard resolution. It is shown here in the $(a,e)$ and 
$(a,i)$ projections. Warmer colors correspond to orbits where NEOs are more likely to be found. The
red lines show borders of the orbital domain where orbits can have close encounters with Earth and
Venus. The black line corresponds to $q=1.3$ au.}
\label{bmodel}
\end{figure}


\clearpage
\begin{figure}
\epsscale{0.8}
\plotone{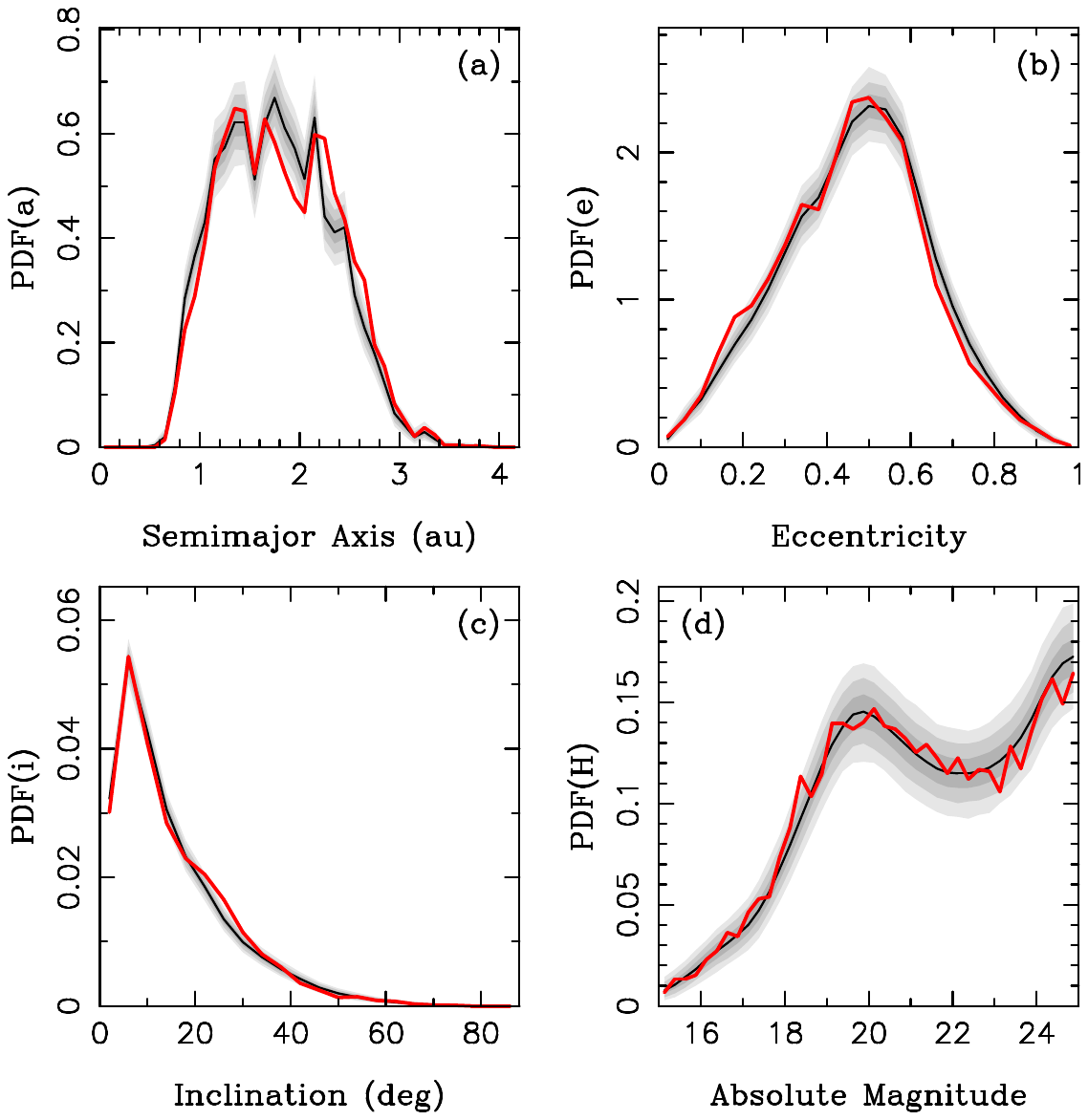}
\caption{The probability density functions (PDFs) of $a$, $e$, $i$, and $H$ from our 
biased base model (black lines) and the CSS2 NEO detections (red lines), both for {\bf bright} 
NEOs with $15<H<25$. The shaded areas are 1$\sigma$ (bold gray), 2$\sigma$ (medium) and 3$\sigma$ 
(light gray) envelopes. We used the best-fit solution (i.e. the one with the maximum likelihood) 
from the base model and generated 30,000 random samples with 8365 NEOs each (the sample size 
identical to the number of CSS2's NEOs with $15<H<25$). The samples were biased and binned with the 
standard binning. We identified envelopes containing 68.3\% (1$\sigma$), 95.5\% 
(2$\sigma$) and 99.7\% (3$\sigma$) of samples and plotted them here.}
\label{bright}
\end{figure}

\clearpage
\begin{figure}
\epsscale{0.8}
\plotone{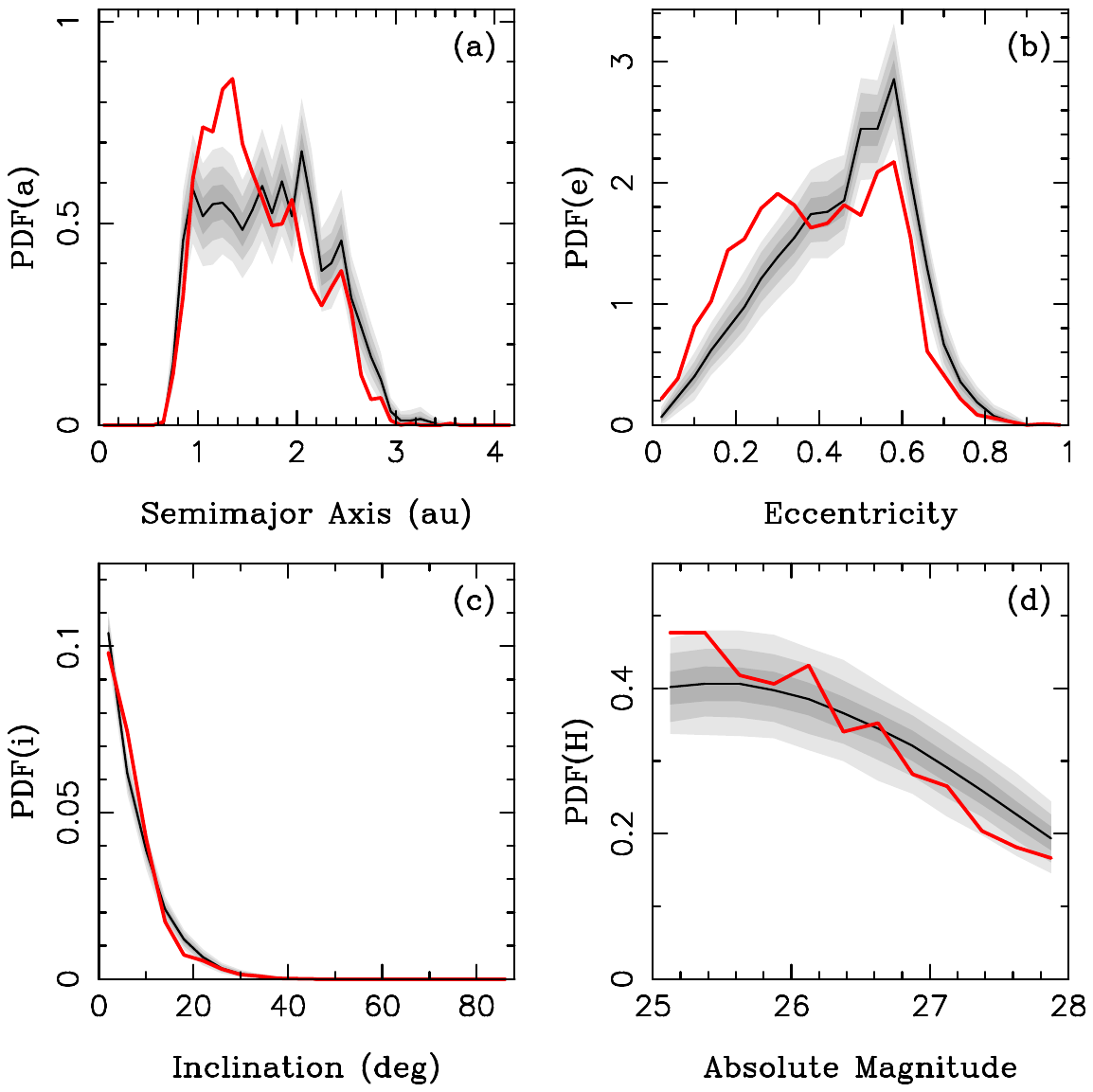}
\caption{The probability density functions (PDFs) of $a$, $e$, $i$, and $H$ from our 
biased base model (black lines) and the CSS2 NEO detections (red lines), both for {\bf faint}
NEOs with $25<H<28$. The shaded areas are 1$\sigma$ (bold gray), 2$\sigma$ (medium) and 3$\sigma$ 
(light gray) envelopes. We used the best-fit solution (i.e. the one with the maximum likelihood) 
from the base model and generated 30,000 random samples with 3003 NEOs each (the sample size 
identical to the number of CSS2's NEOs with $25<H<28$). The samples were biased and binned with the 
standard binning. We identified envelopes containing 68.3\% (1$\sigma$), 95.5\% 
(2$\sigma$) and 99.7\% (3$\sigma$) of samples and plotted them here.}
\label{faint}
\end{figure}

\clearpage
\begin{figure}
\epsscale{1.5}
\hspace*{-4cm}\plotone{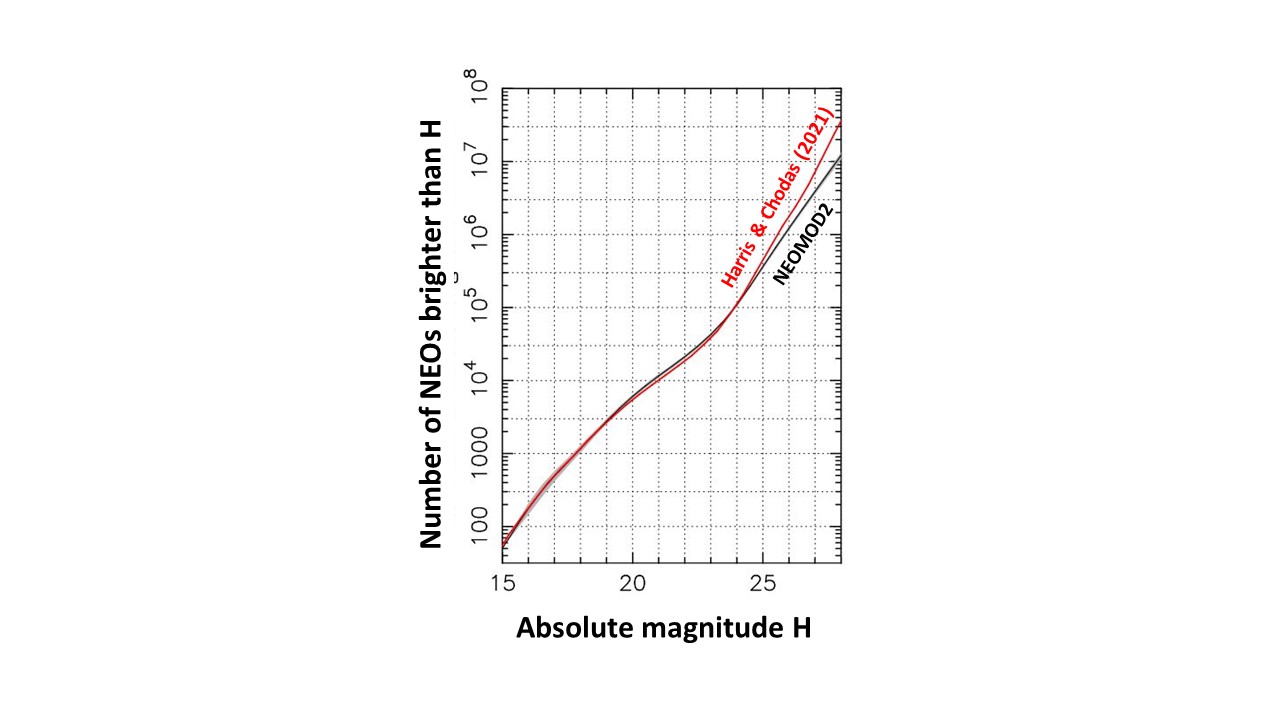}
\caption{The intrinsic (debiased) absolute magnitude distribution of NEOs from our base model
(black line is the median) is compared to the magnitude distribution from Harris \& Chodas (2021)
(red line). The gray area is the 3$\sigma$ envelope obtained from the posterior distribution 
computed by \texttt{MultiNest}. It contains -- by definition -- 99.7\% of our base model 
posteriors.}
\label{harris}
\end{figure}

\clearpage
\begin{figure}
\epsscale{0.6}
\plotone{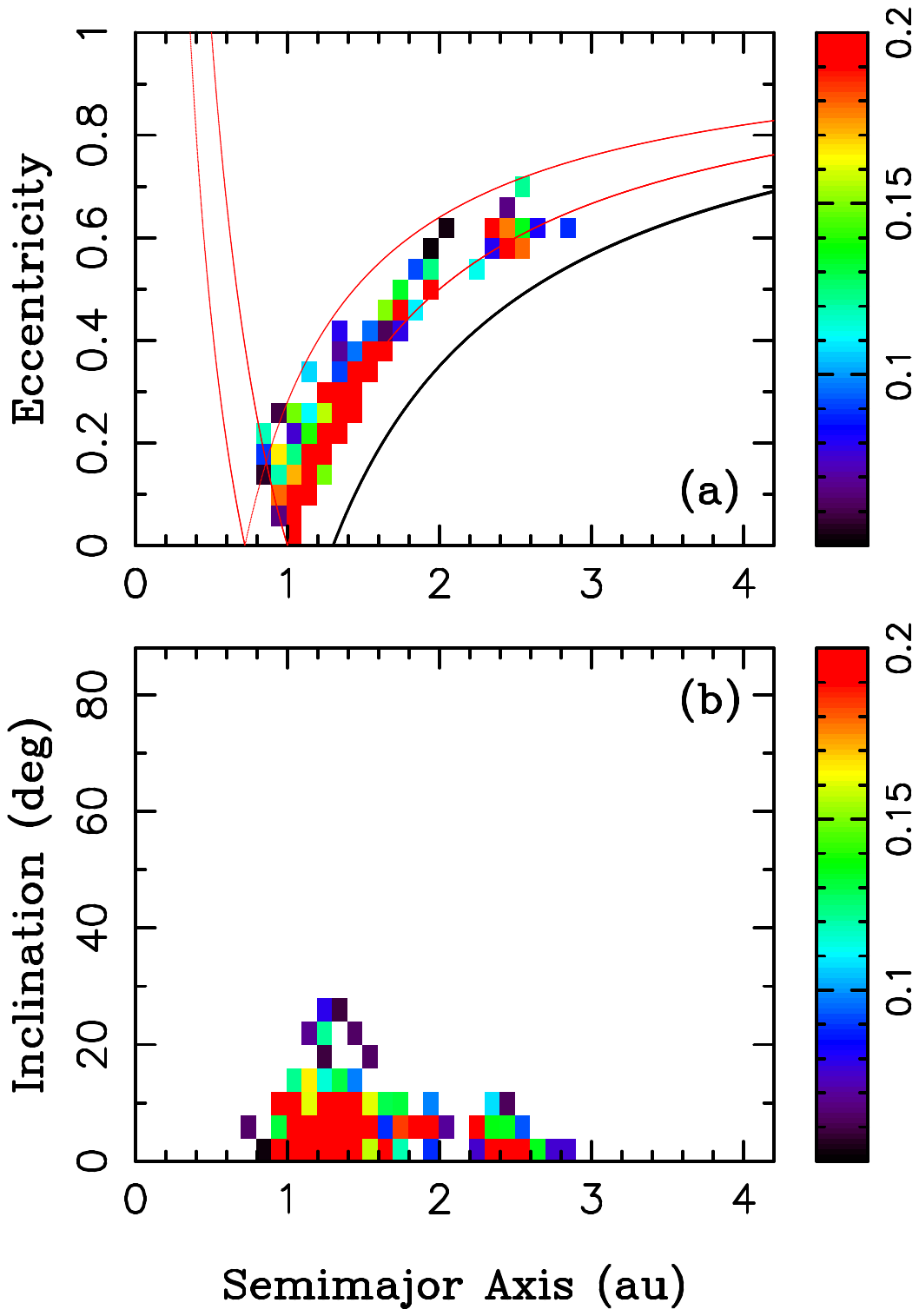}
\caption{The excess of $25<H<28$ NEOs detected by new CSS relative to our base model for $25<H<28$ 
(Sect. 4). We binned NEOs detected by CSS2 with the standard binning (Paper I), subtracted the number of NEOs 
predicted in each bin by our biased best fit model, ${\cal M}_{\rm b}$, and normalized it by ${\cal M}_{\rm b}$.
The red color shows that the largest excess, roughly 20--30\%, happens for $1<a<1.6$ au, $q \simeq 1$ au 
and $i\lesssim 10^\circ$. The red lines show borders of the orbital domain where orbits can have close
encounters with Earth and Venus. The black line corresponds to $q=1.3$ au.}
\label{excess}
\end{figure}

\clearpage
\begin{figure}
\epsscale{0.8}
\plotone{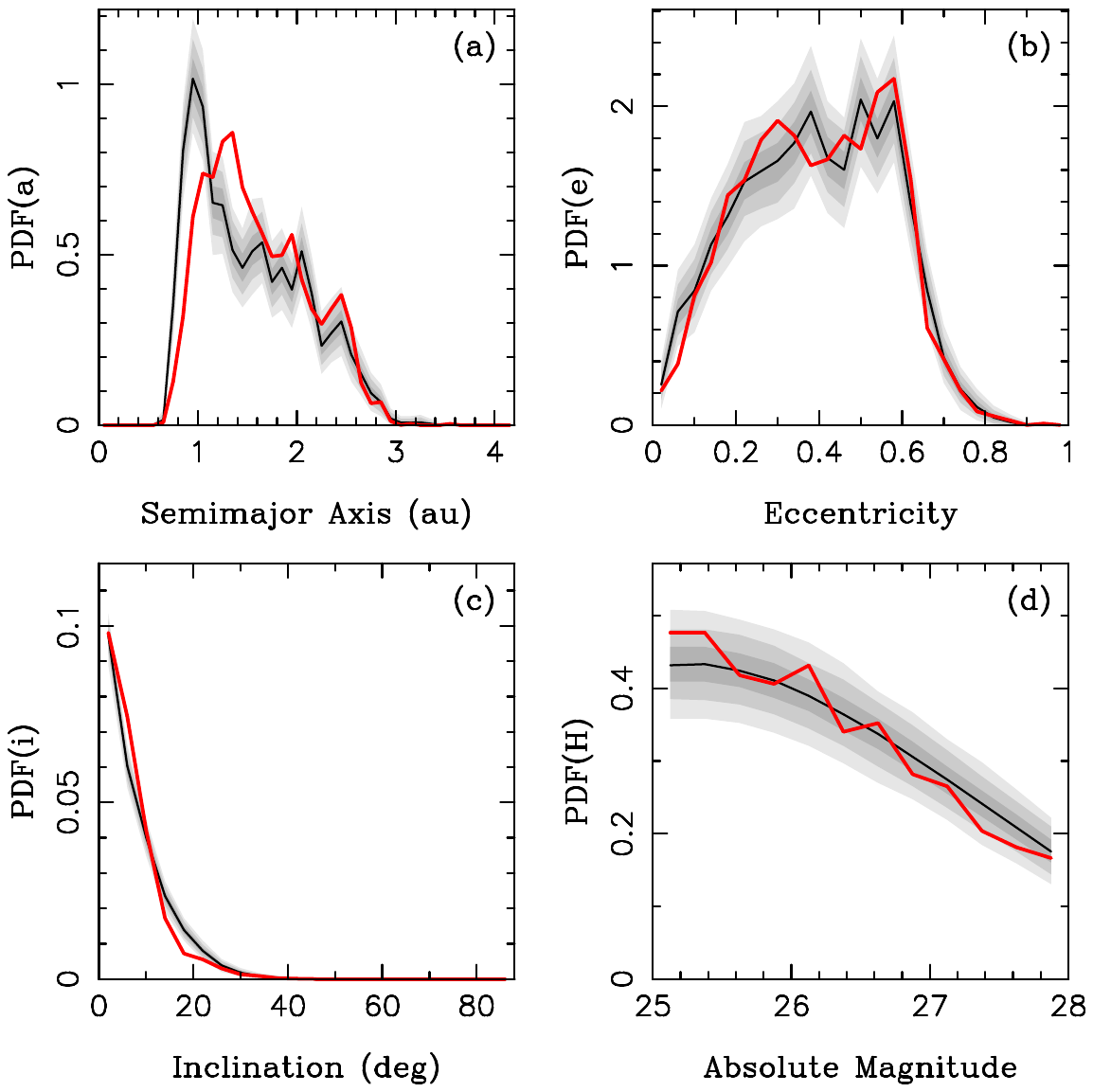}
\caption{The probability density functions (PDFs) of $a$, $e$, $i$, and $H$ from our 
biased based model with tidal disruption (black lines) and the CSS2 NEO detections (red lines),
both for faint NEOs with $25<H<28$. The shaded areas are 1$\sigma$ (bold gray), 2$\sigma$ (medium) 
and 3$\sigma$ (light gray) envelopes. We used the best-fit solution (i.e. the one with the maximum 
likelihood) from the model with tidal disruption and generated 30,000 random samples with 3003 NEOs each (the sample size 
identical to the number of CSS2's NEOs with $25<H<28$). The samples were biased and binned with the 
standard binning. We identified envelopes containing 68.3\% (1$\sigma$), 95.5\% 
(2$\sigma$) and 99.7\% (3$\sigma$) of samples and plotted them here.}
\label{tidal}
\end{figure}


\begin{figure}
\epsscale{0.7}
\epsscale{1.5}
\hspace*{-4cm}\plotone{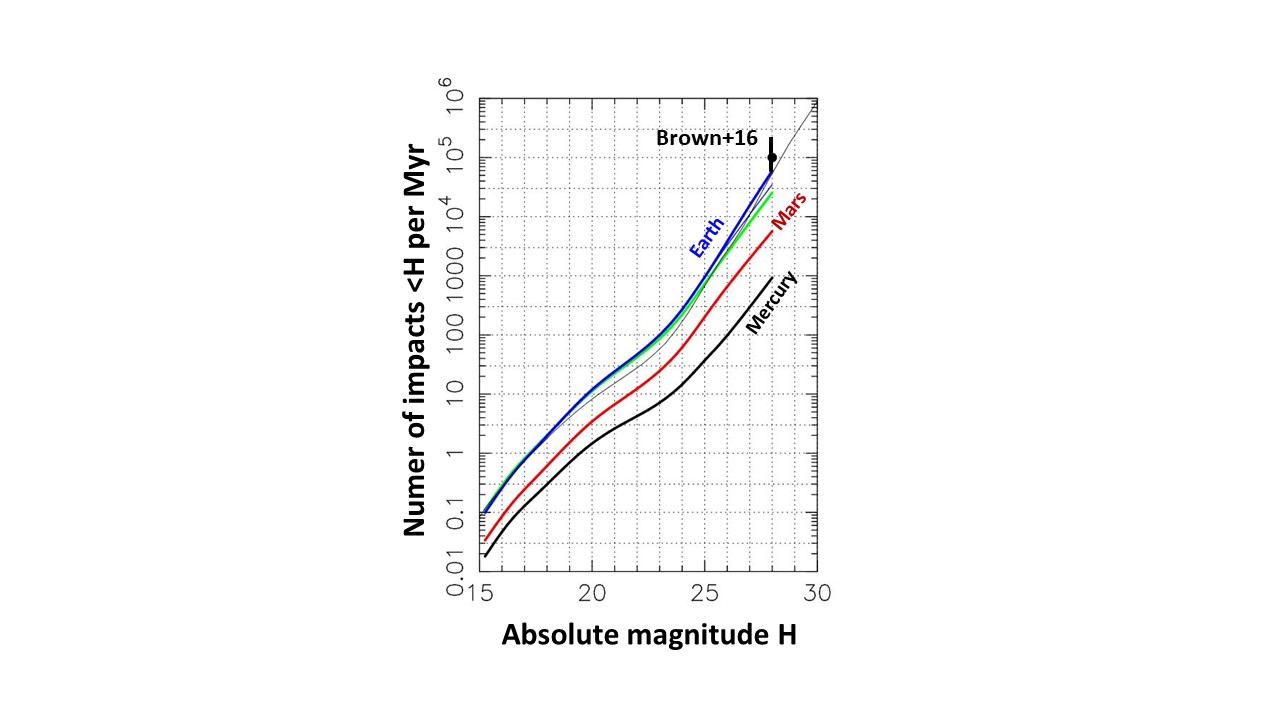}
\caption{The impact flux on the terrestrial planets for our base model with tidal disruption.
The black, green, blue and red lines show the impact flux for Mercury, Venus, Earth and Mars from Eq. (\ref{fimp}).
With a 30\% contribution of tidal disruptions at $H=28$, the mean time between impacts of $D>10$ m NEOs is 
$\simeq 17$ years (was $\simeq 30$ years in the base model without tidal disruption; thin blue line), 
which is more consistent with bolide and infrasound observations (Brown et al. 2002, 2013; black dot).
The thin black line is the NEO magnitude distribution from Harris \& Chodas (2021) scaled 
with the fixed impact probability ($1.5 \times 10^{-3}$ Myr$^{-1}$; see the main text).}
\label{impacts}
\end{figure}

\clearpage
\begin{figure}
\epsscale{0.5}
\plotone{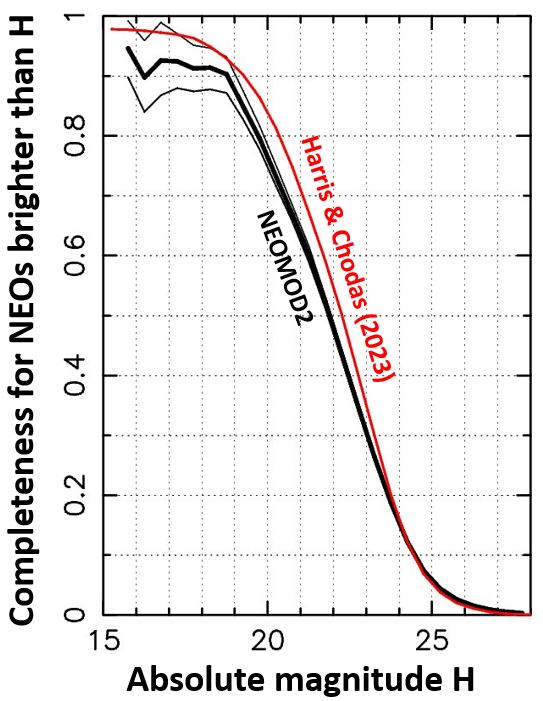}
\caption{The estimated completeness of the NEO population from our base model (thick black line; the 
two thin lines show 1 sigma uncertainty; Table 3) is compared to the completeness estimated in Harris \& Chodas 
(2023) (red line). For $H<19$, the completeness given in Harris \& Chodas (2023) is consistent with
1 sigma envelope of our results. The redetection method may provide a more accurate completeness 
estimate for these bright NEOs. Our model indicates slightly lower completeness than Harris \& Chodas 
(2023) for $19<H<24$, and slightly higher completeness than Harris \& Chodas (2023) for $H>24$.}
\label{compl}
\end{figure}

\end{document}